\documentclass[preprint,prd,aps,nofootinbib]{revtex4}
\usepackage{graphicx}
\usepackage{diagbox}
\usepackage{float}
\usepackage{mathrsfs}
\usepackage{color}
\usepackage{slashed}
\begin{document}
\title{Radiative transitions of $\chi_{_{cJ}}\to\psi\gamma$ and $\chi_{_{bJ}}\to\Upsilon\gamma$}
\author{Su-Yan Pei$^{1,2,3}$\footnote{peisuya@163.com},
Wei Li$^{1,2,3}$\footnote{watliwei@163.com}, Tianhong Wang$^{4}$\footnote{thwang@hit.edu.cn},
Guo-Li Wang$^{1,2,3}$\footnote{wgl@hbu.edu.cn, corresponding author}}

\affiliation{$^1$ Department of Physics, Hebei University, Baoding 071002, China
\nonumber\\
$^{2}$ Hebei Key Laboratory of High-precision Computation and Application of Quantum Field Theory, Baoding 071002, China
\nonumber\\
$^{3}$ Hebei Research Center of the Basic Discipline for Computational Physics, Baoding 071002, China
\nonumber\\
$^{4}$ School of Physics, Harbin Institute of Technology, Harbin 150001, China}
\begin{abstract}
In the framework of instantaneous Bethe-Salpeter equation, according to the $J ^ {PC}$ of quarkonia, we find that their wave functions all contain multiple partial waves, rather than pure waves.
In the radiative electromagnetic transitions $\chi_{_{cJ}}$$\rightarrow$$\gamma\psi$ and $\chi_{_{bJ}}$$\rightarrow$$\gamma\Upsilon$ ($J=0,1,2$), the main wave of quarkonium gives the non-relativistic contribution, while other waves provide the relativistic corrections.
Our results indicate that the relativistic effect of charmonium, especially highly excited states, is significant. Such as
the relativistic effects of $\chi_{_{cJ}}(2P)\to\gamma\psi(1S)$ ($J=0,1,2$) are $\{49.7\%,~30.9\%,~37.5\%\}$, much larger than the corresponding $\{17.8\%,~7.08\%,~12.9\%\}$ of $\chi_{_{bJ}}(2P)\rightarrow\gamma\Upsilon(1S)$.
The decay of $\chi_{_{cJ}}(2P)\to\gamma\psi$ can be used to distinguish between $\chi_{_{c0}}(3860)$ and $\chi_{_{c0}}(3915)$, which particle is the charmonium $\chi_{_{c0}}(2P)$.
Although our result of $\chi_{_{c1}}(3872)$$\rightarrow$$\gamma\psi(2S)$  is consistent with data, but the one of $\chi_{_{c1}}(3872)$$\rightarrow$$\gamma\psi(1S)$ is much larger than data, so whether $\chi_{_{c1}}(3872)$ is the conventional $\chi_{_{c1}}(2P)$ remains an open question.
The undiscovered $\Upsilon(1D)$ and $\Upsilon(2D)$ have large production rates in decays of $\chi_{_{b0}}(2P)\rightarrow\gamma\Upsilon(1D)$ and $\chi_{_{bJ}}(3P)\rightarrow\gamma\Upsilon(2D)$ ($J=0,1$), respectively. To search for $\chi_{_{bJ}}(3P)$ $(J=0,1,2)$, the most competitive channels are the decays $\chi_{_{bJ}}(3P)\rightarrow\gamma\Upsilon(3S)$. And the best way to find $\chi_{_{b2}}(1F)$ is to search for the decay of $\chi_{_{b2}}(1F)\rightarrow\gamma\Upsilon(1D)$.

\end{abstract}
\maketitle

\section{Introduction}
Heavy quarkonium has been a highly active research field in high energy physics since they were discovered \cite{Eichten:1974af,article,Gaiser:1985ix}. In the past two decades, significant progress has been made in the study of heavy quarkonia. For example, in experiments, charmonia have been widely studied by the collaborations of BESIII
\cite{BESIII2010-2,BESIII2017gcu,BESIII2019-1},
Belle \cite{Belle:2003nnu,Belle:2010,Belle:2012},
BaBar \cite{BaBar1,BaBar2,BaBar3},
CDF \cite{CDF2,CDF:2006ocq,CDF3} and
D0 \cite{D01,D02}; Bottomonia by BABAR \cite{BaBar2,BaBar:2006fjg}, Belle \cite{Belle1,Belle2} and CLEO \cite{CLEO:2010xuh,CLEO2008pkg}, etc.
In theory, various aspects of heavy quarkonia, including the mass spectra, decays, productions, and so on, have been studied \cite{Duan2020tsx,Barnes_2005,Segovia:2018qzb,Li:2009zu,Wang:2010ej,Chang:2010kj,Ebert:2002pp,Anwar:2018yqm,Brambilla2020ojz,Baranov2015yea,SPDMIXING,guo,Wang:2022dfd}. Interested readers can find more details in the review articles \cite{Brambilla2019esw,zsl,Chen2016qju,Brambilla2010cs,Eichten2007qx}.

There are still many heavy quarkonia that have not yet been detected and we know very little about them, such as the highly excited states $\chi_{_{c2}}(3P)$, $\chi_{_{b0}}(3P)$, $\chi_{_{c2}}(1F)$, $\Upsilon(1D)$, $\Upsilon(2D)$ and $\chi_{_{b2}}(1F)$, etc. There are also some quarkonia that have been experimentally found, but their properties are still controversial.
$X(3872)$ \cite{Belle:2003nnu}, also known as $\chi_{_{c1}}(3872)$, is a typical representative of this type of particle. It was discovered  twenty years ago, but we still do not fully understand its internal structure. It may be a conventional charmonium $\chi_{_{c1}}(2P)$, tetraquark state, molecular, or a mixture of them. To reveal its nature, one important factor is the ratio of its electromagnetic decays, $\frac{Br(X(3872)\to \psi(2S)+\gamma)}{Br(X(3872)\to J/\psi+\gamma)}$, currently the experimental data is $2.6\pm0.6\%$ \cite{PDG}.
Another example is $\chi_{_{c0}}(3915)$ and $\chi_{_{c0}}(3860)$, $\chi_{_{c0}}(3915)$ was first discovered \cite{39151,3915}, and it was proposed as the radial excitation $\chi_{_{c0}}(2P)$. But this assignment is challenged because   $\chi_{_{c0}}(3860)$ is discovered \cite{3860}, the later is also has the quantum number of $0^{++}$ and being the candidate of $\chi_{_{c0}}(2P)$. In terms of bottomonia, the experiment has not yet detected the total widths of all the $\chi_{_{bJ}}$. {While there are the branching ratios of radiative transitions $\chi_{_{bJ}}(nP)\to\gamma+\Upsilon(mS)$ $(n\geq m)$ in experiments \cite{PDG}, and for particles without strong decay allowed by OZI-rule, the branching ratios are sizable, for examples, $Br(\chi_{b1}(1P)\to\gamma\Upsilon(1S))=(35.2\pm2.0)\%$, $Br(\chi_{b2}(1P)\to\gamma\Upsilon(1S))=(18.0\pm1.0)\%$,
and $Br(\chi_{b1}(2P)\to\gamma\Upsilon(2S))=(18.1\pm1.9)\%$. So studying these electromagnetic transitions is important for determining the full widths of $\chi_{_{bJ}}(nP)$ ($n\leq3$).}

So in this paper, using the Bethe-Salpeter method, we study the electromagnetic radiative transitions
$\chi_{_{cJ}}\to\gamma+\psi$ and $\chi_{_{bJ}}\to\gamma+\Upsilon$ ($J=0,1,2$), where
the quantum number of the initial quarkonium is $0^{++}$, $1^{++}$ or $2^{++}$, and those of the final quarkonium is $1^{--}$. In detail, the following processes
$\chi_{_{cJ}}(nP)\to\gamma+\psi(mS)$, $\chi_{_{bJ}}(nP)\to\gamma+\Upsilon(mS)$ $(n=1,2,3;~n\geq m)$;
$\chi_{_{cJ}}(nP)\to\gamma+\psi(mD)$, $\chi_{_{bJ}}(nP)\to\gamma+\Upsilon(mD)$ $(n=2,3;~n>m)$;
$\chi_{_{c2}}(1F)\to\gamma+\psi(mS)$ ($m=1,2$); $\chi_{_{b2}}(1F)\to\gamma+\Upsilon(mS)$ ($m=1,2,3$);  $\chi_{_{c2}}(1F)\to\gamma+\psi(1D)$; and $\chi_{_{b2}}(1F)\to\gamma+\Upsilon(1D)$, etc., will be studied.

{In previous study \cite{vsquare}, the square of the charm quark velocity inside the charmonium is calculated, and $\langle v^2_{J/\psi}\rangle\approx\langle v^2_{\eta_c(1S)}\rangle\approx0.25$, $\langle v^2_{\psi(2S)}\rangle\approx\langle v^2_{\eta_c(2S)}\rangle\approx0.34$, $\langle v^2_{\psi(3S)}\rangle\approx\langle v^2_{\eta_c(3S)}\rangle\approx0.41$, $\langle v^2_{\chi_{cJ}(1P)}\rangle\approx0.30$, $\langle v^2_{\chi_{cJ}(2P)}\rangle\approx0.38$ and  $\langle v^2_{\chi_{cJ}(3P)}\rangle\approx0.45$ are obtained. These results indicate that the excited states of charmonium, especially the highly excitations, have significant relativistic corrections. So  it is necessary to choose a relativistic method to study them.} The Bethe-Salpeter (BS) equation \cite{Salpeter:1951sz} is a relativistic dynamic equation for bound state, but it is too complicated to be solved, we have to make approximations to apply it. Salpeter equation \cite{ret8} is the instantaneous version of BS equation, which is suitable for heavy mesons. We will solve the complete Salpeter equation without further approximation, therefore, we can solve wave functions with up to eight unknown variables \cite{ret1-}, so we can have a relatively complete relativistic wave function. Due to the fact that the BS or Salpeter equation is an integral equation, not a differential equation, so in this method, the relativistic correction in the wave function is equivalent to calculating all orders of the expansion method.
For example,  in non-relativistic method, charmonium radiative transition $^3D_2\to {^3}P_1 +\gamma$  is an $E_1$ decay, while in our method this transition  includes the results of $E_1+M_2+E_3+...$ \cite{Li:2022qhg}.
Further, the transition amplitude is described as the overlapping integral over the relativistic wave functions of the initial and final mesons, where there are no more approximations except for the instantaneous approximation \cite{gauge}. Therefore, this BS method is very effective in dealing with heavy meson problems and our theoretical results are in good agreement with experimental data \cite{Liu:2022kvs,ret13}.

In literature, a particle, for example, $\chi_{_{c0}}(1P)$ is usually considered to be a pure $P$-wave quarkonium. But in our method, since the wave function is a relatively complete relativistic one, it contains more information, and shows that this particle is not a pure $P$-wave, but also contains a small amount of $S$-wave component. Specifically, $\psi(3770)$ is not a pure $D$-wave meson, not only a $S-D$ mixing state, but also a $S-P-D$ mixing state. It is interesting to make a detailed study of the behaviors of different partial waves. So in this paper, after  presenting the results of electromagnetic decays,
$\chi_{_{cJ}}$$\rightarrow$$\gamma\psi$ and $\chi_{_{bJ}}$$\rightarrow$$\gamma\Upsilon$,  ($J$$=$$0,1,2$), we will choose some processes as examples to discuss in detail the contributions of different partial waves in the initial and final state wave functions.

This paper is organized as follows, In Sec. II, we will provide the relativistic wave functions, which include different partial waves, and  show the method how to calculate the transition amplitude. In Sec. III, the results of the electromagnetic decay and the contribution of different partial waves are discussed.
\section{Electromagnetic transition}
\subsection{The positive wave functions}
{ Given that the $1^{--}$ meson $\psi(3770)$ is a $S-D$ mixing state (in our method, it is a $S-P-D$ mixing state) and the $1^+$ meson $D_1(2420)$ is a ${}^1P_1-{}^3P_1$ mixing state, we know that the quantum number $J^{P\{C\}}$ is a more fundamental quantum number than the spin $S$, orbital angular momentum $L$, or the description ${}^{2S+1}L_J$ of a meson. So we provide the general representation of the wave function of $\chi_{_{cJ}}$ and $\chi_{_{bJ}}$ based on $J^{PC}$, rather than ${}^{2S+1}L_J$ \cite{Wang:2022cxy}.}

{ Based on the quantum number $J^{PC}=0^{++}$, the general representation of the wave function for a $0^{++}$ state under instantaneous approximation is written as \cite{ret01+},
\begin{equation}\label{0++}
\varphi_{_{0^{++}}}(q_{\bot})=a_{1}\slashed{q}_{\bot}+a_{2}\slashed{P}\slashed{q}_{\bot}/M+a_{3}M,
\end{equation}
where $M$ is the mass, $P$ is the total momentum of the meson, respectively. $q$ is the internal relative momentum between the quark and the anti-quark, which can be divided into $q_{_{\parallel}}$ and $q_{_{\bot}}$, where $q_{_{\parallel}}^{\mu}=(P\cdot q/M^{2})P^{\mu}$, $q_{_{\bot}}^{\mu}=q^{\mu}-q_{_{\parallel}}^{\mu}$, in the center of mass system of the meson, they become to $q_0$ and $\vec{q}$, respectively. Because we chose the instantaneous approximation, that is $P\cdot q=P\cdot q_{_{\bot}}=0$, the wave function does not contain the term of $P\cdot q$.
The radial part of the wave function $a_i$ is function of $-q_{_\bot}^2$ ($=\vec{q}^2$ in the center of mass system), and it is unknown.  Its numerical solution is obtained by solving the Salpeter equation satisfied by the $0^{++}$ state \cite{ret01+}, and we have the relation $a_3=\frac{a_1q_{_\bot}^2}{Mm}$, where $m$ is the quark mass.

By using the formulas $\gamma_0\varphi_{_{P'}}({q'})\gamma_0$ and $C\varphi^{T}_{_P}({-q})C^{-1}$, where $P'=(P_0,-\vec P)$, $q'=(q_0,-\vec q)$ and $C=\gamma_2\gamma_0$,
it can be checked that each term in Eq.(\ref{0++}) has positive parity and positive charge conjugate parity. }

{In the BS method, instead of the wave function itself, the commonly used one is the positive energy wave function. And the positive energy wave function of the $0^{++}$ ($\chi_{_{c0}}$ or $\chi_{_{b0}}$) meson can be written as, }
\begin{eqnarray}\label{s0+}
\varphi_{0^{++}}^{++}(q_{\bot})=A_{1}\slashed{q}_{\bot}+A_{2}\frac{\slashed{P}\slashed{q}_{\bot}}{M}+A_{3},
\end{eqnarray}
 $A_i$ ($i=1,2,3$) is the function of $a_i$ (i=1,2):
$$
A_{1}=\frac{1}{2}\left(a_{1}+\frac{m}{w}a_{2}\right),~~A_{2}=\frac{w}{m}A_{1},~~A_{3}=\frac{q_{\bot}^{2}}{m}A_{1},
$$
where $w=\sqrt{m^{2}-q_{\bot}^{2}}$ is the quark energy, respectively.
{ And we have the relation  $\varphi_{_{0^{++}}}(q_{\bot})=\varphi^{++}_{_{0^{++}}}(q_{\bot})+\varphi^{--}_{_{0^{++}}}(q_{\bot})$, where the value of negative energy wave function $\varphi^{--}_{_{0^{++}}}(q_{\bot})$ is much smaller than that of $\varphi^{++}_{_{0^{++}}}(q_{\bot})$.}

In the meson rest frame and in spherical coordinate system, the positive energy wave function (similar to $\varphi_{0^{++}}({q}_{_\bot})$) can be written as,
$$
\varphi_{0^{++}}^{++}({q}_{_\bot})=\displaystyle
 \sqrt{4\pi}\left[ -\frac{|\vec q|}{\sqrt{3}}(A_1
+A_2\gamma^0)( Y_{1-1}\gamma^+ +Y_{11}\gamma^- -Y_{10}\gamma^3)+A_3Y_{00}\right],
$$
where $\gamma^{\pm}=\mp\frac{1}{\sqrt{2}}(\gamma^1\pm \gamma^2)$, and $Y_{Lm}$ is the spherical harmonic function.
So it is easy to see that, $A_1$ and $A_2$ terms are $P$-waves, while $A_3$ term is a $S$-wave. So the $0^{++}$ wave function is not a pure $P$-wave, but a $P$$-$$S$ wave mixture. {In the non-relativistic limit, only $P$ wave term exists, and the wave function is $\varphi_{_{0^{++}}}=a_1(1+\slashed{P}/M)\slashed{q}_{\bot}$. So in Eq.(\ref{0++}) or Eq.(\ref{s0+}), $P$-wave provides the non-relativistic contribution, while $S$-wave gives the relativistic correction.  }
\begin{figure}[htbp]
    \centering
    \includegraphics[scale=0.3]{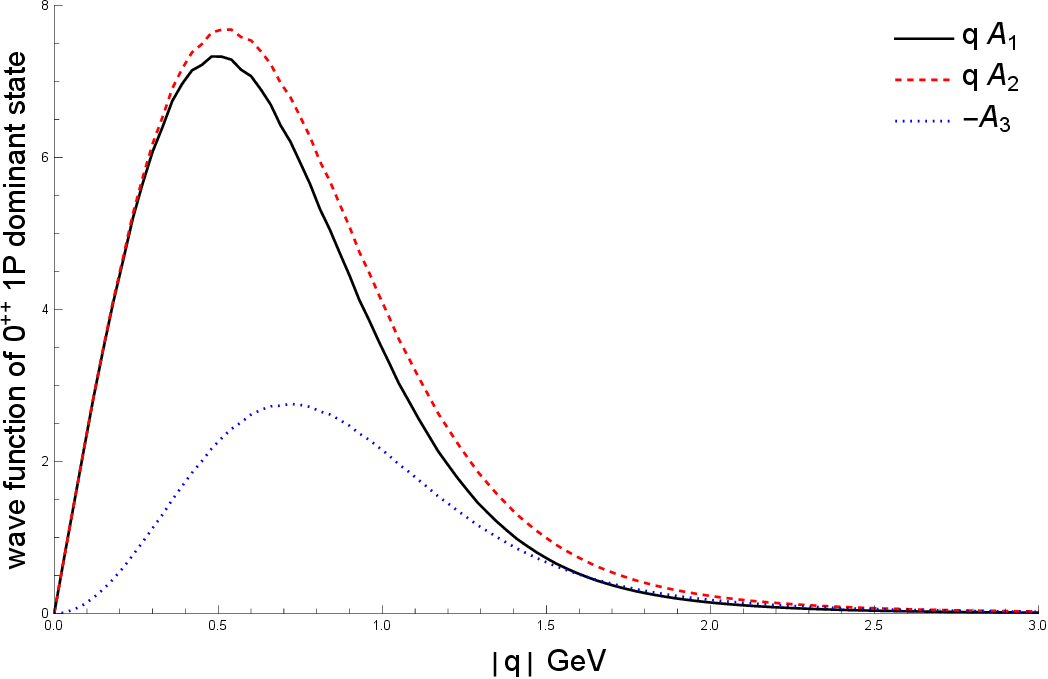}
    \includegraphics[scale=0.3]{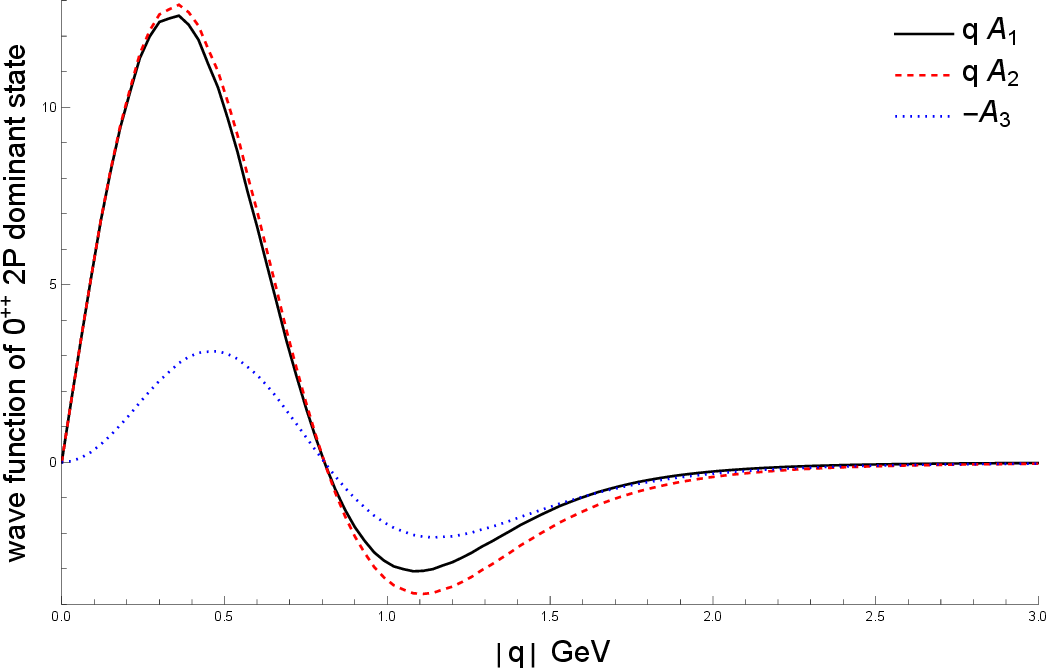}
    \includegraphics[scale=0.3]{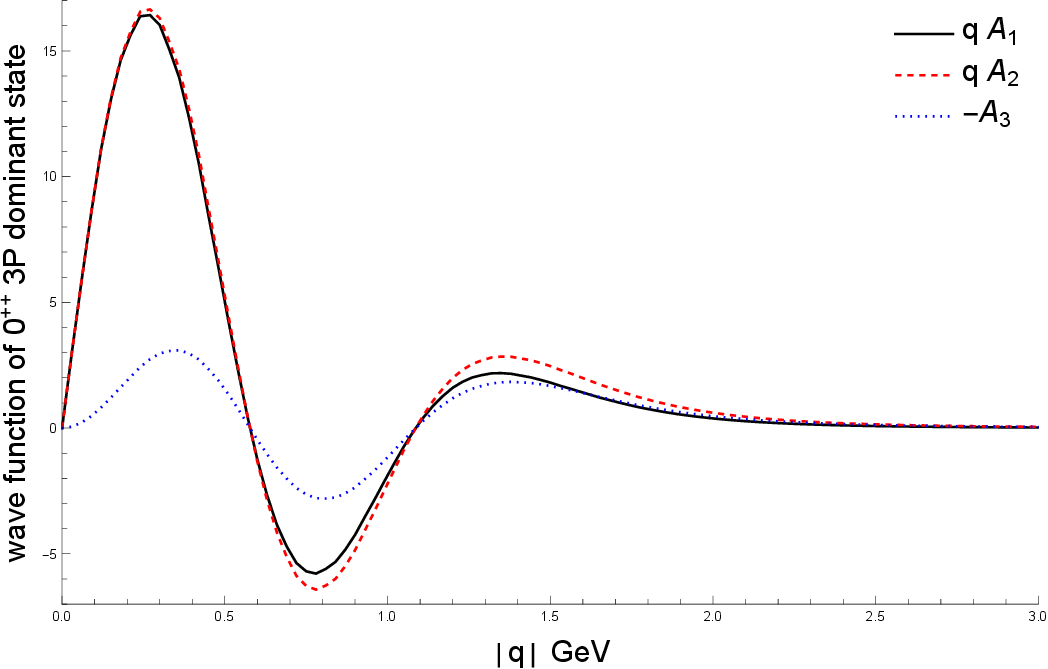}
    \caption{The positive wave functions of the ground state, the first and the second excited states of the $0^{++}$ charmonia, corresponding to $\chi_{c0}(1P)$,  $\chi_{c0}(2P)$ and $\chi_{c0}(3P)$. $q\equiv|q|\equiv|\vec{q}|$. $A_1$ and $A_2$ terms are $P$-waves, and $A_3$ term is a $S$-wave.}\label{pic0++}
    \end{figure}

{
We draw the radial wave functions of the first three solutions of the Salpeter equation for $0^{++}$ charmonium in Figure \ref{pic0++}. The diagrams show us  that $|\vec{q}A_1|\approx |\vec{q}A_2|>|A_3|$, so for $0^{++}$ charmonium, usually, the non-relativistic contribution is dominant, and the relativistic correction provides the secondary contribution. In Fig.\ref{pic0++}, there are no nodes in the wave function of the first diagram, so this solution corresponds to the ground state charmonium $\chi_{c0}(1P)$. The wave function in the second diagram has one node, which corresponds to the first radial excited state $\chi_{c0}(2P)$. The third solution with two nodes corresponds to the second radial excited state $\chi_{c0}(3P)$.

Based on our results, we note that the well-known representations of ${}^{2S+1}L_J$ and $nL$  exactly applies only to the non-relativistic terms, not the relativistic terms.
However, in literature, such as in PDG and the previous paragraph, it is customary to label physical particles with `nL', such as $\eta_c(1S)$, $\eta_c(2S)$ and $\chi_{c0}(1P)$, $\chi_{c0}(2P)$, which correspond to the ground state and first radial excited state of $0^{-+}$ and $0^{++}$ states, respectively.
So in this paper, we still use the familiar symbol $nL$ to label the physics states, where $n$ is the principal quantum number, and $L$ is the orbital angular momentum of the meson, as did in the previous paragraph.
}

The positive energy wave function for a $1^{++}$ meson is written as \cite{ret01+}:
\begin{eqnarray}\label{s1+}
\varphi_{1^{++}}^{++}(q_{\bot})=i\epsilon_{\eta\nu\alpha\beta}P^{\nu}q_{\bot}^{\alpha}\epsilon^{\beta}\gamma^{\eta}\left(B_{1}\frac{1}{M}+B_{2} \frac{\slashed{P}}{M^2}+B_{3}\frac{\slashed{P}\slashed{q}_{\bot}}{M^2}\right),
\end{eqnarray} 
where $\varepsilon_{\eta\nu\alpha\beta}$ is the Levi-Civita tensor, $\epsilon^{\beta}$ is the polarization vector of the meson. 
Eq.(\ref{s1+}) shows that the $1^{++}$ wave function is $P$$-$$D$ wave mixture, that is, the non-relativistic $B_1$ and $B_2$ terms are $P$ waves, while the relativistic $B_3$ term is $D$ wave. { We show the diagrams of positive wave functions of first three solutions of the Salpeter equation for $1^{++}$ states, and they corresponding to $\chi_{c1}(1P)$,  $\chi_{c0}(2P)$ and $\chi_{c0}(3P)$ states.}

\begin{figure}[htbp]
    \centering
    \includegraphics[scale=0.3]{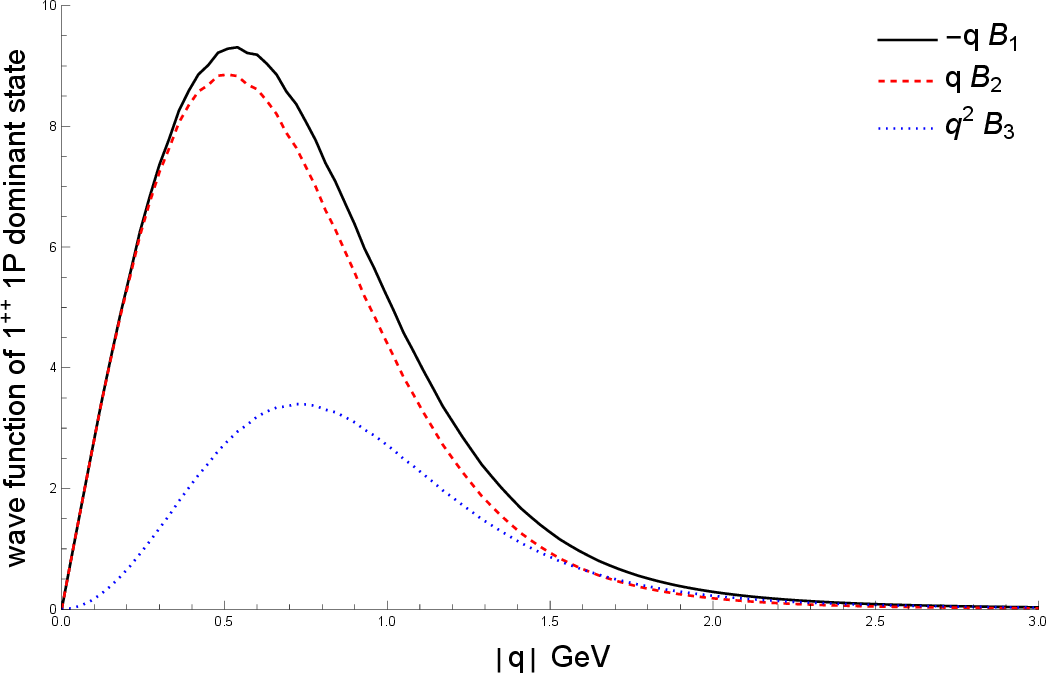}
    \includegraphics[scale=0.3]{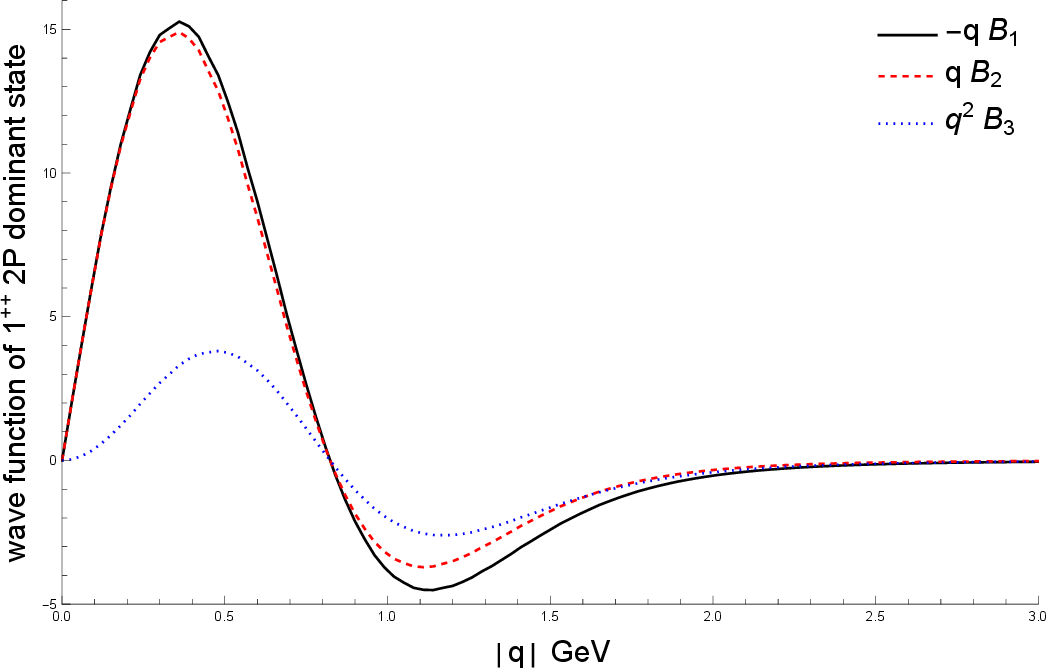}
    \includegraphics[scale=0.3]{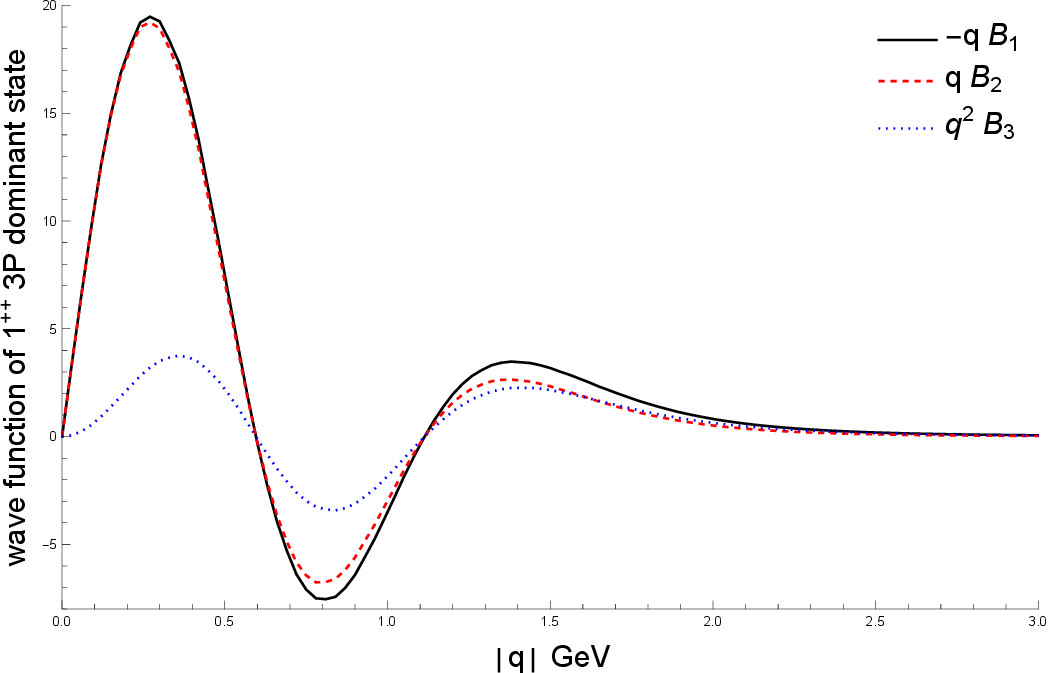}
    \caption{The positive wave functions of the $1^{++}$ charmonia  $\chi_{c1}(1P)$,  $\chi_{c0}(2P)$ and $\chi_{c0}(3P)$. $B_1$ and $B_2$ terms are $P$-waves, and $B_3$ term is a $D$-wave.}\label{pic1++}
    \end{figure}

The positive energy wave function for a $2^{++}$ state is written as \cite{ret2+}:
\begin{eqnarray}\label{2+}
\varphi_{2^{++}}^{++}(q_{\bot})=&&\epsilon_{\rho\sigma}q_{\bot}^{\rho}q_{\bot}^{\sigma}\left(C_{1}+\frac{\slashed{P}}{M}C_{2} +\frac{\slashed{q}_{\bot}}{M}C_{3}+\frac{\slashed{P}\slashed{q}_{\bot}}{M^{2}}C_{4}\right)
\nonumber\\
&&+M\epsilon_{\rho\sigma}\gamma^{\rho}q_{\bot}^{\sigma}\left(C_{5}+\frac{\slashed{P}}{M}C_{6}+\frac{\slashed{P}\slashed{q}_{\bot}}{M^{2}}C_{7}\right),
\end{eqnarray}
where $\epsilon_{\rho\sigma}$ is the polarization tensor of the $2^{++}$ meson. 

\begin{figure}[htbp]
    \centering
    \includegraphics[scale=0.35]{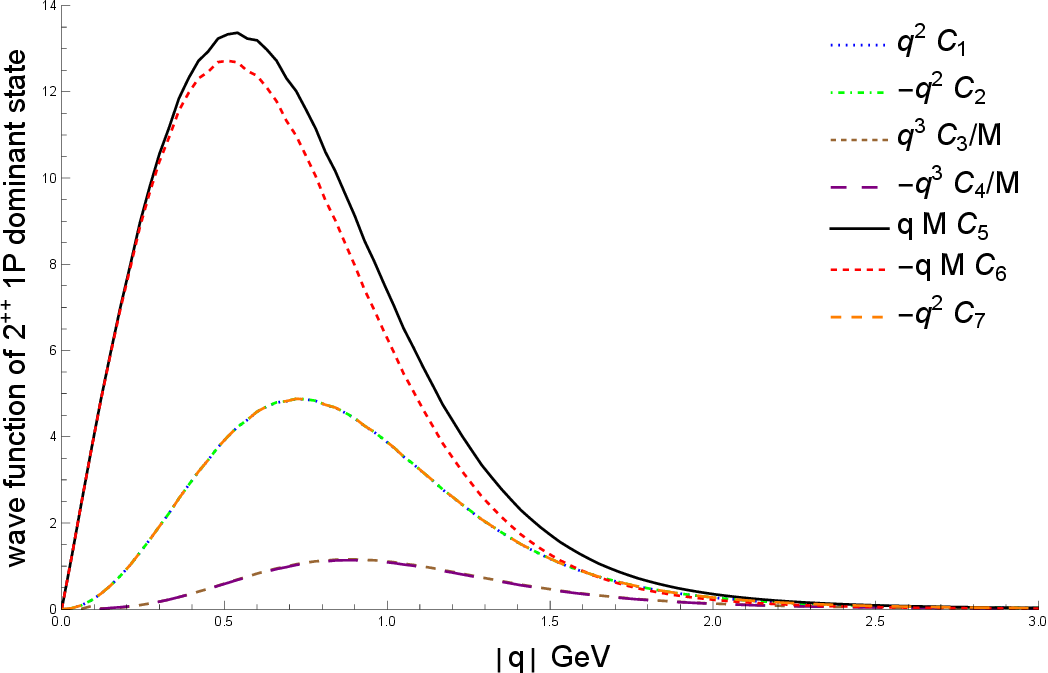}
    \includegraphics[scale=0.35]{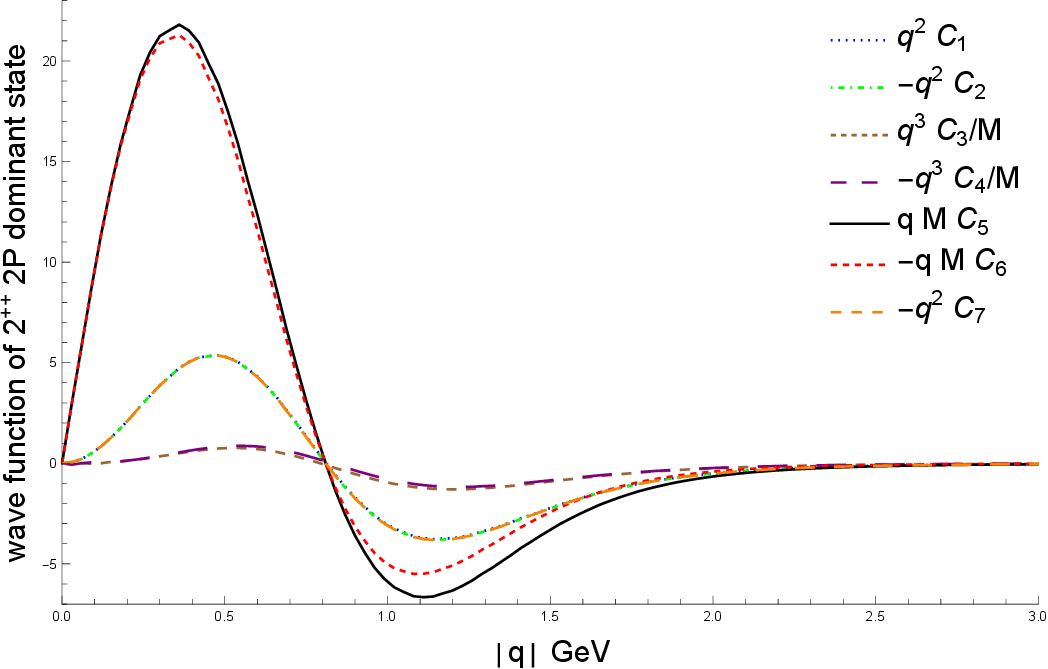}
    \includegraphics[scale=0.35]{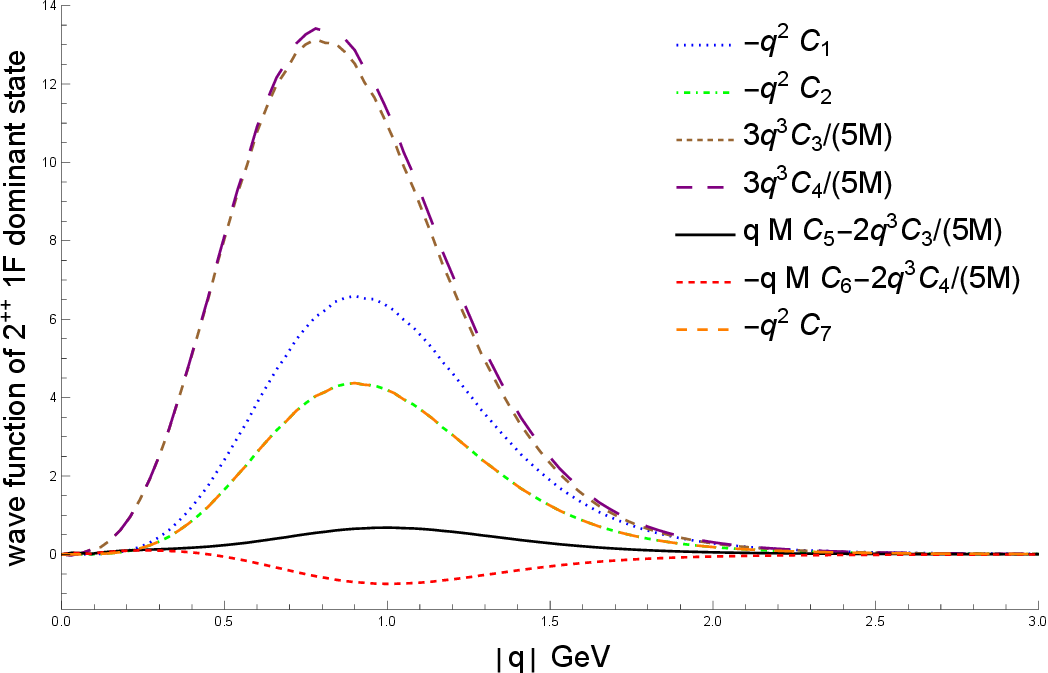}
    \includegraphics[scale=0.35]{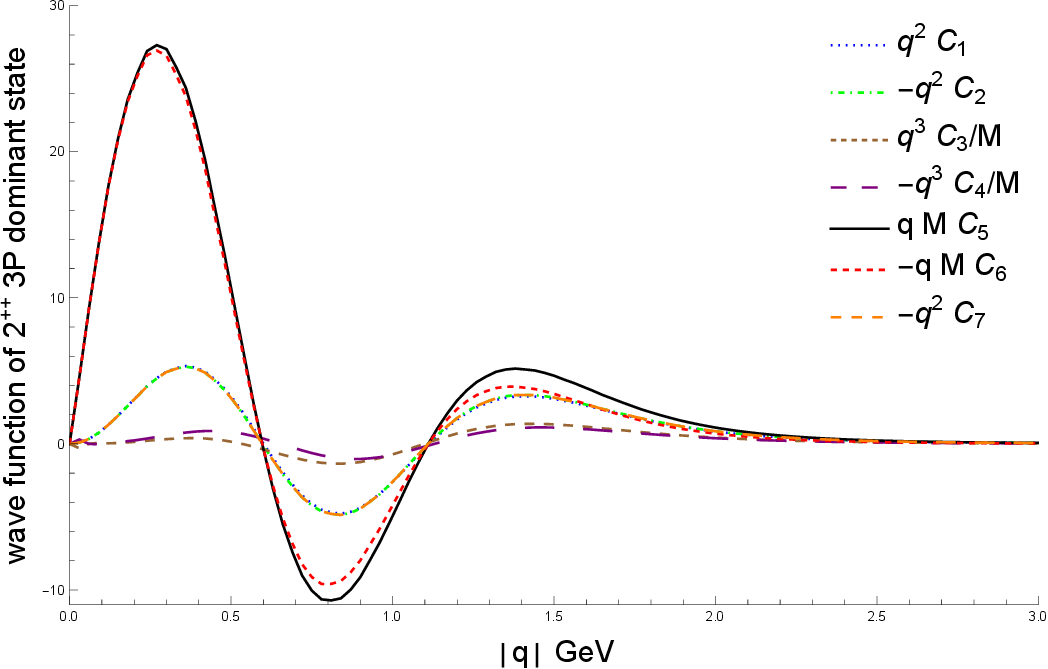}
    \caption{The positive wave functions of the first four  solutions of Salpeter equation for $2^{++}$ state, which corresponding to charmonia $\chi_{c2}(1P)$, $\chi_{c2}(2P)$, $\chi_{c2}(1F)$ and $\chi_{c2}(3P)$.}\label{pic2++}
    \end{figure}

The partial waves in $2^{++}$ wave function is complicated, this tensor meson is a $P$$-$$D$$-$$F$ mixing state. In Eq.(\ref{2+}), $C_5$ and $C_6$ terms are $P$ waves, $C_1$, $C_2$ and $C_7$ terms are $D$ waves, while $C_3$ and $C_4$ terms are $P$$-$$F$ mixture. We can represent Eq.(\ref{2+}) with the spherical harmonic function $Y_{_{Lm}}$, the term with $L$$=$$1$ is the $P$-wave, the one with $L$$=$$3$ is $F$-wave. Thus the $F$ waves in Eq. (\ref{2+}) are
\begin{equation} \left(\epsilon_{\rho\sigma}q_{_\bot}^{\rho}q_{_\bot}^{\sigma}\slashed{q}_{\bot}
-\frac{2}{5}q_{_\bot}^{2}\epsilon_{\rho\sigma}\gamma^{\rho}q_{_\bot}^{\sigma}\right)\left(\frac{1}{M}C_{3}-\frac{\slashed{P}}{M^{2}}C_{4}\right),
\end{equation}
and the complete $P$ waves are
\begin{equation}
\epsilon_{\rho\sigma}\gamma^{\rho}q_{_\bot}^{\sigma}\left[M\left(C_{5}+\frac{\slashed{P}}{M}C_{6}\right)
+\frac{2}{5}q_{_\bot}^{2}\left(\frac{1}{M}C_{3}-\frac{\slashed{P}}{M^{2}}C_{4}\right)\right].
\end{equation}

{In Figure \ref{pic2++}, we show the positive wave functions of the first four solutions of the complete Salpeter equation with the wave function form Eq.(\ref{2+}) as input. We can see that, unlike the cases of $0^{++}$ and $1^{++}$, there are two types of solutions.  One type of solution includes the first, the second, and the fourth solutions, they are dominated by $P$-waves, the $C_5$ and $C_6$ terms, which provide the non-relativistic contribution; The $D$-waves, $C_1$, $C_2$ and $C_7$ terms, are sizable and will contribute to the relativistic corrections; While the $P-F$ mixtures, $C_3$ and $C_4$ terms, are very small and can be ignored. The second type includes the third and the fifth solutions. In these solutions, the $C_3$, $C_4$, $C_5$ and $C_6$ terms are all large. If we distinguish different waves, as shown in the third diagram of Fig.\ref{pic2++}, it can be seen that the wave function is dominated by $F$-waves, $\frac{3|\vec q|^3C_3}{5M}$ and $\frac{3|\vec q|^3C_4}{5M}$, and they will provide the non-relativistic contribution; $D$-wave $C_1$, $C_2$ and $C_7$ terms are sizable and contribute to the relativistic corrections; Due to the cancellations between the $P$-waves from $C_3$, $C_4$ and from $C_5$ and $C_6$, so the total $P$-waves, $|\vec q|MC_5-\frac{2|\vec q|^3C_3}{5M}$ and $|\vec q|MC_6+\frac{2|\vec q|^3C_4}{5M}$ are very small and can be ignored.}

\begin{figure}[htbp]
    \centering
    \includegraphics[scale=0.3]{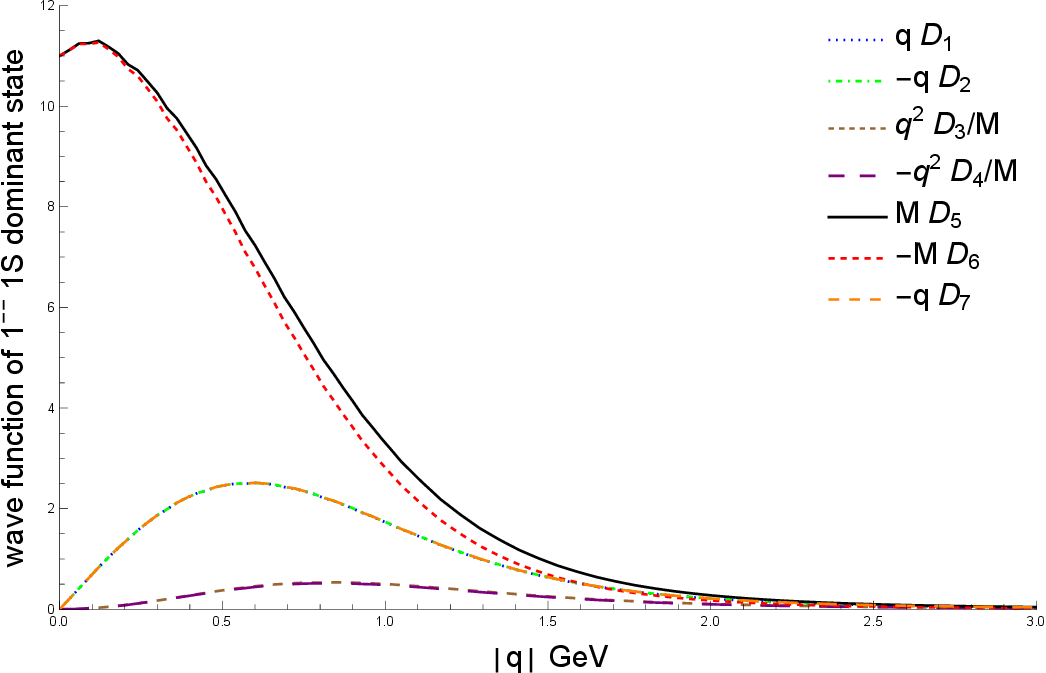}
    \includegraphics[scale=0.3]{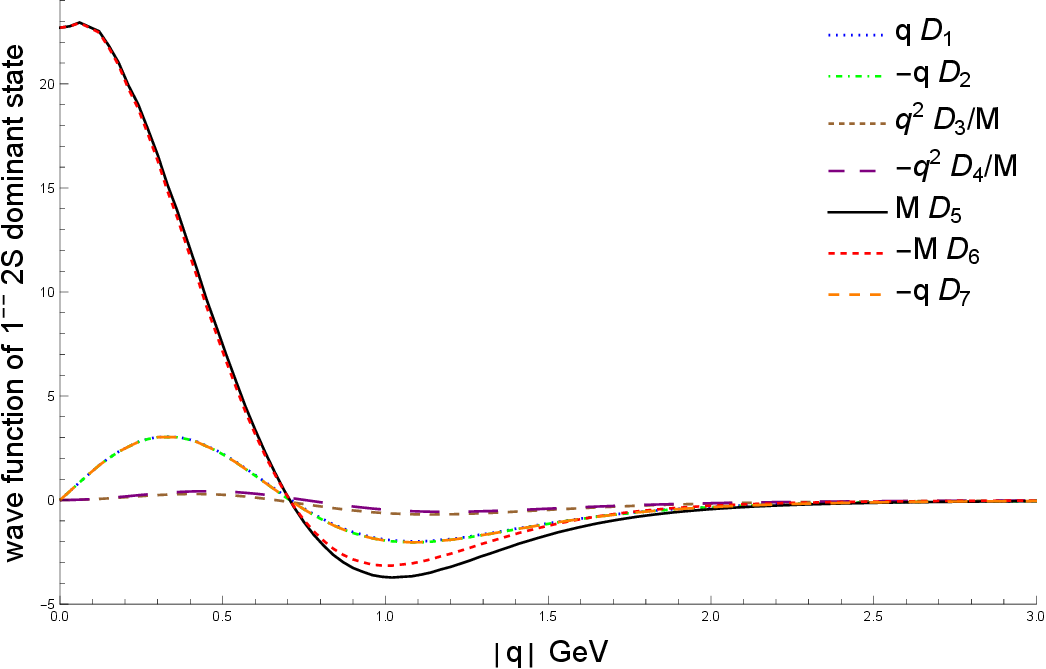}
    \includegraphics[scale=0.3]{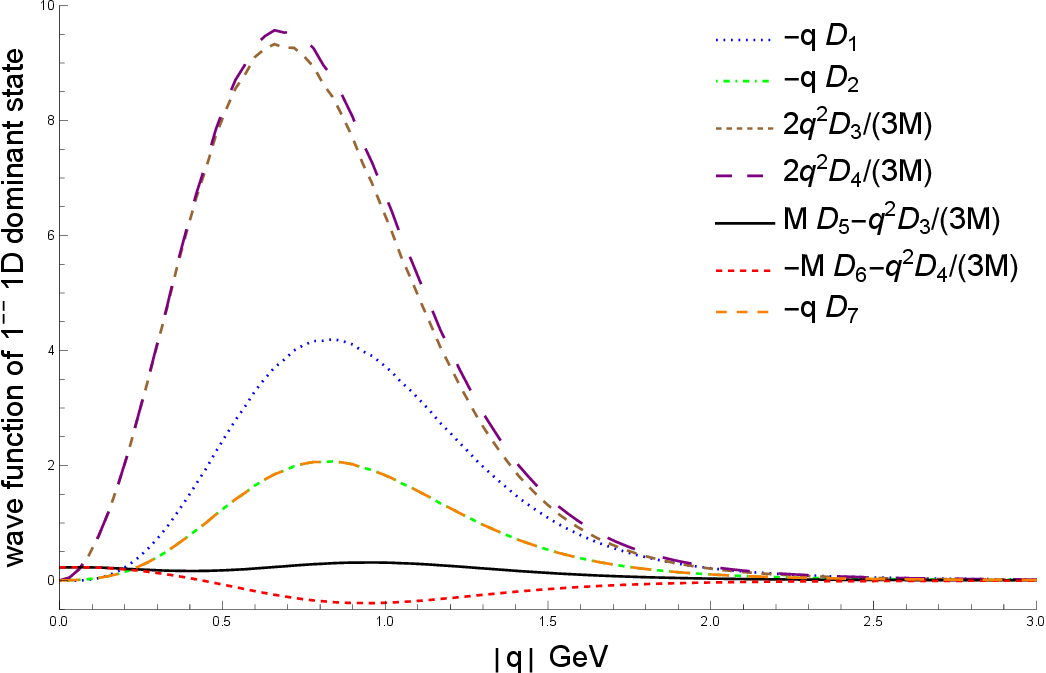}
    \includegraphics[scale=0.3]{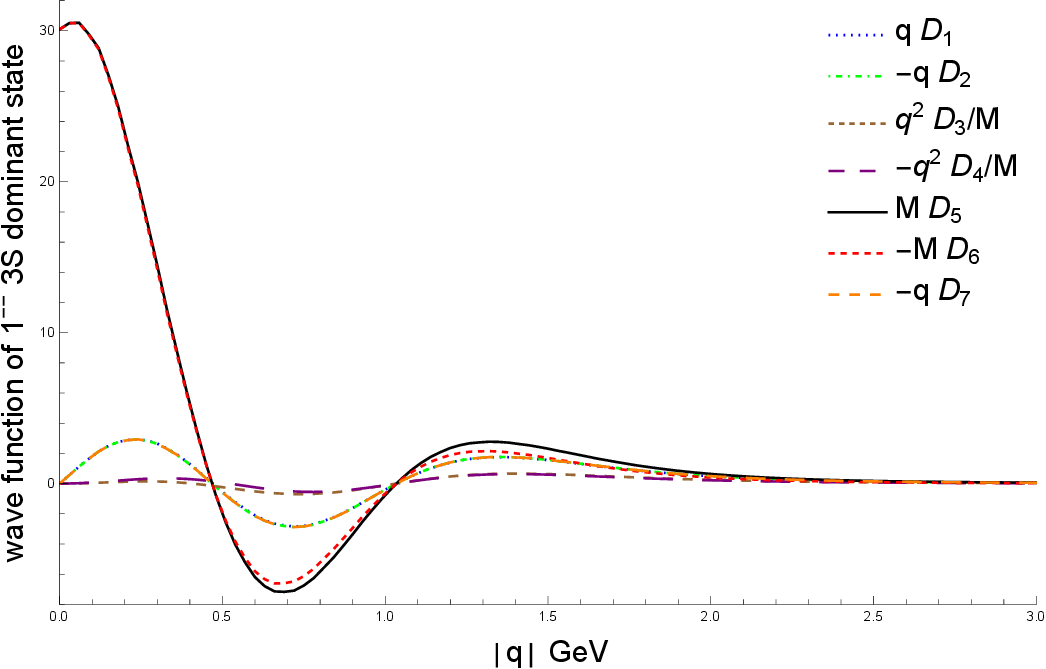}
    \includegraphics[scale=0.3]{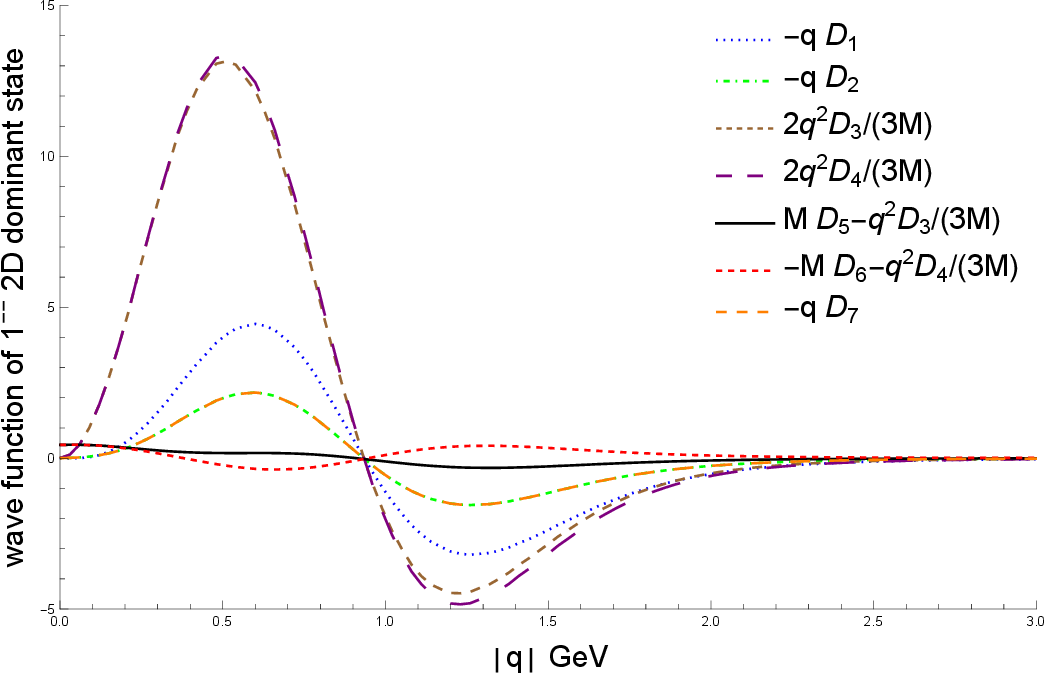}
    \caption{The positive wave functions of the first five  solutions of Salpeter equation for $1^{--}$ state, which corresponding to charmonia $\psi(1S)$, $\psi(2S)$, $\psi(1D)$, $\psi(3S)$ and $\psi(2D)$.}\label{pic1-}
    \end{figure}

The positive energy wave function of $1^{--}$ state is \cite{ret1-}:
\begin{eqnarray}\label{1-}
\varphi_{1^{--}}^{++}(q_{\bot})=&&q_{\bot}.\epsilon\left(D_{1}+\frac{\slashed{P}}{M}D_{2}+\frac{\slashed{q}_{\bot}}{M}D_{3}+\frac{\slashed{P}\slashed{q}_{\bot}}{M^{2}}D_{4}\right)
\nonumber\\
&&+M\slashed{\epsilon}\left(D_{5}+\frac{\slashed{P}}{M}D_{6}+\frac{\slashed{P}\slashed{q}_{\bot}}{M^{2}}D_{7}\right),
\end{eqnarray}
where $\epsilon$ is the polarization vector of the meson. 
$1^{--}$ meson is a $S$$-$$P$$-$$D$ mixing state. The terms including $D_5$ and $D_6$ are $S$ waves, $D_1$, $D_2$ and $D_7$ terms are $P$ waves, while $D_3$ and $D_4$ terms are $S-D$ mixture. The $D$ waves in Eq.(\ref{1-}) are
\begin{equation}
\left(\epsilon\cdot q_{_\bot}\slashed{q}_{\bot}
-\frac{1}{3}q_{_\bot}^{2}\slashed{\epsilon}\right)
\left(\frac{1}{M}D_{3}-\frac{\slashed{P}}{M^{2}}D_{4}\right),
\end{equation}
and the complete $S$ waves are
\begin{equation}\label{Swave}
M\slashed{\epsilon}\left(D_{5}+\frac{\slashed{P}}{M}D_{6}\right)+\frac{1}{3}q_{_\bot}^{2}\slashed{\epsilon}\left(\frac{1}{M}D_{3}-\frac{\slashed{P}}{M^{2}}D_{4}\right).
\end{equation}
{Similar to the $2^{++}$ cases, see Figure \ref{pic1-}, there are two categories of solutions for the $1^{--}$ wave functions.  First one includes the first, the second, and the fourth solutions, they are dominated by $S$-waves, $D_5$ and $D_6$ terms, and provide the non-relativistic contributions; $P$-waves, $D_1$, $D_2$ and $D_7$ terms, are sizable and give the relativistic corrections; The $S-D$ mixture, $D_3$ and $D_4$ terms, are tiny and can be ignored. Second category includes the third and the fifth solutions, they are dominated by $D$-waves from $D_3$ and $D_4$ terms; $P$-waves are the relativistic correction terms; $S$-waves, $MD_5-\frac{|\vec q|^2D_3}{3M}$ and $MD_6+\frac{|\vec q|^2D_4}{3M}$, are very small and can be ignored.}

\subsection{Interaction kernel}
In our method, since the wave function is fully relativistic, the interaction kernel should be a non-relativistic one. Generally, the relativistic correction can be reflected in the wave function or in the interaction potential. This is because the relativistic potential is obtained by expanding the relativistic Dirac spinors, and the remaining basic components are constructed as the non-relativistic wave function. If without expansion, the relativistic Dirac spinors directly become to the relativistic wave function. The method with a non-relativistic wave function and a relativistic potential, is good at the calculation of mass spectrum. While the method with a relativistic wave function but a non-relativistic potential, is good at not only the mass spectrum, but also the calculating of transition processes.

{So in our calculation, a simple non-relativistic potential, Cornell potential \cite{Eichten:1974af,Eichten:1978tg}, which is a linear scalar confinement potential plus a single gluon exchange vector Coulomb potential, is chosen,
\begin{equation}
V(r)=V_{s}(r)+ \gamma_0\otimes\gamma^0 V_{v}(r)= \lambda r+V_0-\gamma_0\otimes\gamma^0\frac{4}{3}\frac{\alpha_s}{r},\label{potential}
\end{equation}
where $\lambda$ is the string tension, $V_0$ is a free constant appearing in potential model to fit data, $\alpha_s$ is the
running coupling constant.
We solve the Salpeter equation in momentum space, when Eq.(\ref{potential}) is transformed into momentum space, divergence occurs at $\vec{q}\rightarrow 0$. To avoid this, in coordinate space, the kernel is changed to \cite{laermann,Huang:1996bk}
\begin{equation}
V(r)= \frac{\lambda}{\alpha}(1-e^{-\alpha
r})+V_0-\gamma_0\otimes\gamma^0\frac{4}{3}\frac{\alpha_s}{r}e^{-\alpha
r}.\label{potential1}
\end{equation}
It can be seen that when parameter $\alpha\rightarrow 0$, Eq.(\ref{potential1}) becomes Eq.(\ref{potential}), so a small value of $\alpha=0.06$ GeV is used in this paper. Ref. \cite{Huang:1996bk} pointed out that with parameter $e^{-\alpha
r}$ such interaction potential also incorporate the color screening effect. And the screened potential can reduce the mass splitting of charmonium, making it better consistent with experimental data \cite{Li:2009zu}.
With these changes, in momentum space, the kernel is
\begin{eqnarray}
V_{s}(\vec{q})=&&-(\frac{\lambda}{\alpha}+V_{0})\delta^{3}(\vec{q})+\frac{\lambda}{\pi^{2}}\frac{1}{(\vec{q}^{2}+\alpha^{2})^{2}},
\nonumber\\
V_{v}(\vec{q})=&&-\frac{2}{3\pi^{2}}\frac{\alpha_{s}(\vec{q})}{(\vec{q}^{2}+\alpha^{2})},
\end{eqnarray}
where $e=2.7183$, and the running coupling constant is
$$\alpha_{s}(\vec{q})=\frac{12\pi}{33-2N_f}\frac{1}{log(e+\frac{\vec{q}^{2}}{\Lambda^2_{QCD}})},$$
where $N_f=3,4$ for charmonium and bottomonium. }

\subsection{Electromagnetic Radiative Decays}
In this section, we show how to calculate the electromagnetic radiative decays of $\chi_{_{cJ}}$ and $\chi_{_{bJ}}$ ($J=1,2,3$) to $1^{--}$ quarkonium. The transition amplitude for one photon radiative decay of a quarkonium is written as:
\begin{eqnarray}
&&T=<P_{_{f}}\epsilon_{_f},k\epsilon_{_0}|S|P\epsilon>=(2\pi)^{4}ee_{_{q}}\delta^{4}(P_{_{f}}+k-P)\epsilon_{_{0}\mu}\mathcal{M}^{\mu},
\end{eqnarray}
where $P$, $P_{f}$, and $k$ are the momenta of initial meson, the final meson, and the photon, respectively; $\epsilon$, $\epsilon_{_f}$, and $\epsilon_{_0}$ are the polarization vectors of the initial quarkonium, final quarkonium and the photon, respectively; $ee_{q}$ is the charge of the charm quark or bottom quark in unit of $e$. For the matrix element $\mathcal{M}^{\mu}$, it can be expressed as an overlapping integral of the initial and final mesons positive wave functions \cite{gauge},
\begin{eqnarray}\label{trans}
\mathcal{M}^{\mu}=\int\frac{d^{3}q_{_{\bot}}}{(2\pi)^{3}}\left\{Tr\left[\frac{\slashed{P}}{M}\overline{\varphi}_{_{P_{_{f}}}}^{++}(q_{_{1\bot}})\gamma^{\mu}\varphi_{_{P}}^{++}(q_{_{\bot}})
-\overline{\varphi}_{_{P_{_{f}}}}^{++}(q_{_{2\bot}})\frac{\slashed{P}}{M}\varphi_{_{P}}^{++}(q_{_{\bot}})\gamma^{\mu}\right]\right\},
\end{eqnarray}
where $\varphi_{_{P}}^{++}$ and $\overline{\varphi}_{_{P_{_{f}}}}^{++}=\gamma_0{\varphi}_{_{P_{_{f}}}}^{\dagger}\gamma_0$ are the positive energy wave functions of initial and final mesons. In this electromagnetic decay process, the emitted photon can be released by the quark in the quarkonium, corresponding to the first term in Eq.(\ref{trans}), or by anti-quark, corresponding to the second term. In the final quarkonium, the relative momentum $q_{_1}$ or $q_{_2}$, is related to the initial relative momentum, $q_{_{1\bot}}=q_{_{\bot}}+\frac{1}{2}P_{f_{\bot}}$ or $q_{_{2\bot}}=q_{_{\bot}}-\frac{1}{2}P_{f_{\bot}}$.

After integrating over the internal relative momentum, the transition matrix elements can be expressed as functions of form factors,
\begin{eqnarray}\label{shape}
\mathcal{M}^{\mu}_{_{0^{++}\rightarrow1^{--}}}=&&P^{\mu}(\epsilon_{_{f}}\cdot P)t_{_{1}}+\epsilon_{_{f}}^{\mu}t_{_{2}},
\nonumber\\
\mathcal{M}^{\mu}_{_{1^{++}\rightarrow1^{--}}}=&&P^{\mu}\epsilon^{\epsilon \epsilon_{_{f}} P P_{_{f}}}x_{_{1}}
+(\epsilon_{_{f}}\cdot P)\epsilon^{\mu \epsilon P P_{_{f}}}x_{_{2}}+\epsilon^{{\mu}\epsilon \epsilon_{_{f}} P}x_{_{3}},
\nonumber\\
\mathcal{M}^{\mu}_{_{2^{++}\rightarrow1^{--}}}=&&P^{\mu}(\epsilon_{_{f}}\cdot P)\epsilon^{P_{_{f}} P_{_{f}}}y_{_{1}}+(\epsilon_{_{f}}\cdot P)\epsilon^{\mu P_{_{f}}}y_{_{2}}+P^{\mu}\epsilon^{\epsilon_{_{f}} P_{_{f}}}y_{_{3}}
+\epsilon^{\mu \epsilon_{_{f}}}y_{_{4}}+\epsilon_{_{f}}^{\mu}\epsilon^{P_{_{f}} P_{_{f}}}y_{_{5}},
\end{eqnarray}
where $t_{_{i}}$, $x_{_{i}}$ and $y_{_{i}}$ are the form factors. The following abbreviations $\epsilon^{P_{_{f}} P_{_{f}}}\equiv \epsilon^{\rho\sigma}P_{_{f \rho}}P_{_{f \sigma}}$, $\epsilon^{\epsilon \epsilon_{_{f}} P P_{_{f}}}=\epsilon^{\eta\nu\alpha\beta}\epsilon_{_\eta} \epsilon_{_{f\nu}} P_{_\alpha} P_{_{f\beta}}$, etc, are used.
In Eq.(\ref{trans}), we have ignored the contribution of the negative energy wave function. Their Lorentz structure is the same as that of positive energy wave function, but their contribution is very small. Without them, the gauge invariance will not be guaranteed automatically  \cite{gauge}.
So the form factors are not all independent, and must meet the Gauge invariance condition $k_{\mu}\mathcal{M}^{\mu}=0$, then we have the following relations,
\begin{eqnarray}
t_{_{2}}=&&(M_{_{f}}^{2}-ME_{_{f}})t_{_{1}},
\nonumber\\
x_{_{3}}=&&(M_{_{f}}^{2}-ME_{_{f}})x_{_{1}},
\nonumber\\
y_{_{1}}=&&(y_{_{3}}-y_{_{5}})/(ME_{_{f}}-M_{_{f}}^{2}),
\nonumber\\
y_{_{4}}=&&(ME_{_{f}}-M_{_{f}}^{2})y_{_{2}},
\end{eqnarray}
where $M_{_{f}}$ and $E_{_{f}}$ are the momentum and energy of the final quarkonium, respectively.

The decay width formula for the electromagnetic radiative process is
\begin{eqnarray}
\Gamma=\frac{|\vec{P_{_f}}|}{8\pi M^{2}}\frac{(ee_q)^2}{2J+1}\sum{|\mathcal{M}|^{2}},
\end{eqnarray}
where $|\vec{P_{_f}}|=(M^2-M^2_{_f})/{2M}$  is the three-dimensional momentum of the final quarkonium.

\section{Numerical Results and discussions}

The parameters used in this paper, $m_b=4.96~\mathrm{GeV}$, $m_c=1.6~\mathrm{GeV}$, $\alpha=0.06$ GeV, $\lambda=0.21$ GeV$^2$, are same as in Ref. \cite{Chang:2010kj}, where we studied the mass spectra of charmonium and bottomonium. Most of our predictions of masses are in good agreement with experimental data, only a few inconsistencies. For example, our prediction of $\chi_{_{c1}}(2P)$ is 3929 MeV, but the experimental data is 3872 MeV, here we treat the $\chi_{_{c1}}(3872)$ as the conventional charmonium. Another is the $\chi_{_{c0}}(2P)$, our prediction is 3837 MeV, while experimental data is 3862 MeV of $\chi_{_{c0}}(3860)$ or 3918 MeV of $\chi_{_{c0}}(3915)$. For these inconsistent particles, we adjust the free parameter $V_0$ to fit data, and give the theoretical calculation. There are some particles that were not found or confirmed in experiment, for these particles, we will use our theoretical predicted masses as follows \cite{Chang:2010kj}:
$
M_{\chi_{_{c0}}(3P)}$=4140 {MeV},  $M_{_{\chi_{_{c1}}(3P)}}$=4228 {MeV},
 $M_{_{\chi_{_{c2}}(3P)}}$=4271 {MeV},
$M_{_{\chi_{_{c2}}(1F)}}$=4037 {MeV},
$M_{_{\Upsilon(1D)}}=10129$ {MeV}, $M_{_{\Upsilon(2D)}}$=10434 {MeV}, $M_{_{\chi_{_{b2}}(1F)}}$=10374 {MeV},
$M_{\chi_{_{b0}}(3P)}$=10524 {MeV},~  $M_{_{\chi_{_{b1}}(3P)}}$=10556 {MeV}, $M_{_{\chi_{_{b2}}(3P)}}$=10561 {MeV}.

\begin{table}[htp]
\begin{center}
\caption{The decay widths of $\chi_{_{cJ}}$$\to$$\gamma\psi$ in unit of keV.} \label{charm}
\begin{tabular}{c c c c c c c c c}
\hline\hline
\textbf{Process} &\textbf{Ours}&NR\cite{Barnes_2005}&GI\cite{Barnes_2005}&\cite{Radford:2007vd}&LP\cite{Deng:2016stx}&SP\cite{Deng:2016stx}&\cite{Li:2009zu}& Ex \cite{PDG}   \\
\hline
$\chi_{_{c0}}(1P)\rightarrow\gamma\psi(1S)$     &132   &152    &114    &139.3 &172   &179   &117     &$151\pm 14$      \\
$\chi_{_{c1}}(1P)\rightarrow\gamma\psi(1S)$     &272   &314    &239    &293.7 &306   &319   &244     &$288\pm 22$      \\
$\chi_{_{c2}}(1P)\rightarrow\gamma\psi(1S)$     &371   &424    &313    &384.1 &284   &292   &309     &$374\pm 27$      \\
\hline
$\chi_{_{c0}}(2P)\rightarrow\gamma\psi(1S)$   &45.1   &56     &1.3    &24.0  &6.1   &2.3   &9.3     &                 \\
$\chi_{_{c0}}(2P)\rightarrow\gamma\psi(2S)$   &55.7   &64     &135    &89.7  &121   &99    &44      &                 \\
$\chi_{_{c0}}(2P)\rightarrow\gamma\psi(3770)$ &12.8   &13     &51     &7.4   &20    &12    &        &                 \\
\hline
$\chi_{_{c1}}(2P)\rightarrow\gamma\psi(1S)$   &39.2   &71     &14     &5.1   &81    &88    &45      &$9.5\pm 6.4$     \\
$\chi_{_{c1}}(2P)\rightarrow\gamma\psi(2S)$   &69.4   &183    &183    &235.8 &139   &155   &60      &$54\pm 33$       \\
$\chi_{_{c1}}(2P)\rightarrow\gamma\psi(3770)$ &4.44   &22     &21     &12.3  &7.9   &9.8   &        &                 \\
\hline
$\chi_{_{c2}}(2P)\rightarrow\gamma\psi(1S)$   &59.5   &81     &53     &36.7  &93    &93    &109     &                 \\
$\chi_{_{c2}}(2P)\rightarrow\gamma\psi(2S)$   &140   &304    &207    &319.4 &135   &150   &100     &                 \\
$\chi_{_{c2}}(2P)\rightarrow\gamma\psi(3770)$ &0.649  &1.9    &1.0    & 0.8  &0.36  &0.46  &        &                 \\
\hline
$\chi_{_{c2}}(1F)\rightarrow\gamma\psi(1S)$     &4.30   &       &       &      &      &      &        &                 \\
$\chi_{_{c2}}(1F)\rightarrow\gamma\psi(2S)$     &0.242  &       &       &      &      &      &        &                 \\
$\chi_{_{c2}}(1F)\rightarrow\gamma\psi(3770)$   &338   &475    &541    &      &      &      &        &                 \\
\hline
$\chi_{_{c0}}(3P)\rightarrow\gamma\psi(1S)$     &11.6   &27     &1.5    &      &0.24  &0.13  &        &                 \\
$\chi_{_{c0}}(3P)\rightarrow\gamma\psi(2S)$     &34.7   &32     &0.045  &      &17    &9.1   &        &                 \\
$\chi_{_{c0}}(3P)\rightarrow\gamma\psi(3770)$   &3.76   &0.037  &9.7    &      &0.27  &0.39  &        &                 \\
$\chi_{_{c0}}(3P)\rightarrow\gamma\psi(4040)$   &19.4   &109    &145    &      &241   &61    &        &                 \\
\hline
$\chi_{_{c1}}(3P)\rightarrow\gamma\psi(1S)$     &13.3   &31     &2.2    &      &50    &45    &        &                 \\
$\chi_{_{c1}}(3P)\rightarrow\gamma\psi(2S)$     &33.8   &45     &8.9    &      &94    &74    &        &                 \\
$\chi_{_{c1}}(3P)\rightarrow\gamma\psi(3770)$   &0.360  &0.014  &0.39   &      &3.2   &2.0   &        &                 \\
$\chi_{_{c1}}(3P)\rightarrow\gamma\psi(4040)$   &130   &303    &181    &      &331   &117   &        &                 \\
$\chi_{_{c1}}(3P)\rightarrow\gamma\psi(4160)$   &0.540  & 19    & 15    &      &8.1   &  0   &        &                 \\
\hline
$\chi_{_{c2}}(3P)\rightarrow\gamma\psi(1S)$     &20.3   &34     &19     &      &61    &51    &        &                 \\
$\chi_{_{c2}}(3P)\rightarrow\gamma\psi(2S)$     &31.2   &55     &30     &      &97    &76    &        &                 \\
$\chi_{_{c2}}(3P)\rightarrow\gamma\psi(3770)$   &0.291  &0.00071&0.001  &      &1.5   &0.79  &        &                 \\
$\chi_{_{c2}}(3P)\rightarrow\gamma\psi(4040)$   &239   &509    &199    &      &281   &114   &        &                 \\
$\chi_{_{c2}}(3P)\rightarrow\gamma\psi(4160)$   &0.0971 & 2.1   &0.77   &      &0.44  &0.004 &        &                 \\
\hline\hline
\end{tabular}
\end{center}
\end{table}

\subsection{The decays of $\chi_{_{cJ}}\to \psi+\gamma$}
In Table \ref{charm}, we show our predictions for the decay widths of $\chi_{_{cJ}}\to \psi+\gamma$, and experimental data and other theoretical results are also listed in the same table for comparison. Where `Ex' means the experimental values from Particle Data Group (PDG) \cite{PDG}, `NR' means the non-relativistic potential model, `GI' means the relativistic Godfrey-Isgur model \cite{Barnes_2005}, `LP' means the linear potential model, and `SP' is the screened potential model \cite{Deng:2016stx}.

\subsubsection{$\chi_{_{cJ}}(1P) \to \psi(1S)+\gamma$ $(J=0,1,2)$}
From Table \ref{charm}, it can be seen that our results are in good agreement with the experimental data. In addition, whether the model is relativistic or non-relativistic, almost all theoretical results listed are comparable to each other and also to the experimental data. There are some reasons for such good agreement results. First, in theoretical calculations, there is no confusion in selecting the mass values, and the used masses of the initial $\chi_{_{cJ}}(1P)$ (or final $\psi(1S)$) are the same. Further, compared with processes containing highly excited states, the meson wave functions in these processes have no nodes, so the overlapping integral of the wave functions is not sensitive to the dependence on the mass of the initial or final meson. Second, as shown below, the relativistic corrections in processes $\chi_{_{cJ}}(1P) \to \psi(1S)+\gamma$ $(J=0,1,2)$ are not large. Third, $\chi_{_{cJ}}(1P) \to \psi(1S)+\gamma$ are all $E_1$ dominated processes, although relativistic methods such as ours include contributions from $M_2$, $E_3$, etc., but the $M_2$ and $E_3$ contributions, to some extent, are equivalent to the relativistic corrections, and are small.

In Tables \ref{charm0++1-} -\ref{charm2P2++1-}, we show the contributions~of~different~partial~waves~to~the~decay~widths.
Where the `whole'  means the result is obtained using the complete wave function, `$S$ wave' means only $S$ wave has contribution and ignoring contributions from other partial waves, etc. In order to  distinguish, we use a `prime' to indicate that the partial wave comes from the final state.
\begin{table}[H]
\begin{center}
\caption{Contributions of different partial waves to decay width (keV) of $\chi_{c0}(1P)$$\to$$\gamma\psi(1S)$.}\label{charm0++1-}
{\begin{tabular}{|c|c|c|c|c|} \hline\hline \diagbox {$0^{++}$}{$1^{--}$}
                     & ~whole~       & ~$S'$ wave~        & $P'$ wave               & ~$D'$ wave~               \\ \hline
  ~whole~               &132            & 146               & 0.636             & ~$4.69\times10^{-3}$~         \\ \hline
  $P$ wave     &~~~~~~~~96.0~~~~~~~~    & 99.9      & ~~~~~~~~0.0434~~~~~~~~    & ~$3.06\times10^{-3}$~         \\ \hline
~~~~~~~$S$ wave~~~~~~~ &2.83    & ~~~~~~~~4.27~~~~~~~~      & 0.347      & ~~~~~~~~$1.74\times10^{-4}$~~~~~~~~  \\ \hline\hline
\end{tabular}}
\end{center}
\end{table}
From the Table \ref{charm0++1-}, we note that $P$ wave in $\chi_{c0}(1P)$ has the dominant contribution, and $S$ wave has small contribution. In state $\psi(1S)$, $S'$ wave provides the largest contribution, $P'$ wave contribution is small, and $D'$ wave has tiny contribution and can be ignored. In the transition of $\chi_{c0}(1P)$$\to$$\gamma\psi(1S)$, the maximum contribution comes from the overlapping integral of $P\times S'$. In non-relativistic cases, $\chi_{c0}(1P)$ only contains $P$ wave and $\psi(1S)$ only contains $S'$ wave from the $D_{_5}$ and $D_{_6}$ terms. Ignoring the $S'$ wave from the $D_{_3}$ and $D_{_4}$ terms, see Eq.(\ref{Swave}), then the non-relativistic contribution is estimated to be $\Gamma_{non-rel}=101$ keV, very close to the result of $P\times S'$, $\Gamma=99.9$ keV. If we represent the complete result with $\Gamma_{rel}=132$ keV, the relativistic effect of this process is estimated to be,
$$\frac{\Gamma_{rel}-\Gamma_{non-rel}}{\Gamma_{rel}}=23.5\%.$$
\begin{table}[H]
\begin{center}
\caption{Contributions of different partial waves to decay width (keV) of $\chi_{c1}(1P)$$\to$$\gamma\psi(1S)$. }\label{charm1++1-}
{\begin{tabular}{|c|c|c|c|c|} \hline \hline\diagbox {$1^{++}$}{$1^{--}$}
                         & ~whole~       & ~$S'$ wave~     & ~$P'$ wave~           & $D'$ wave              \\ \hline
  ~whole~                  & 272             & 245            & 0.617           & $1.17\times10^{-3}$        \\ \hline
~~~~~~$P$ wave~~~~~~~     & 272     & ~~~~~~~~251~~~~~~~~    & 0.413   & ~~~~~~~~$3.54\times10^{-3}$~~~~~~~~\\ \hline
  ~$D$ wave~     & ~~~~~~~~0.0218~~~~~~~~   & 0.0627  & ~~~~~~~~0.0235~~~~~~~~  & $7.57\times10^{-4}$        \\ \hline\hline
\end{tabular}}
\end{center}
\end{table}
Table \ref{charm1++1-} presents the contributions of different partial waves in $\chi_{c1}(1P)$$\to$$\gamma\psi(1S)$, where initial $\chi_{c1}$ contains $P$ wave and $D$ wave, and $P$ wave provides the dominant contribution.
Similar to the case of $\chi_{c0}(1P)$$\to$$\gamma\psi(1S)$, in transition $\chi_{c1}(1P)$$\to$$\gamma\psi(1S)$, the maximum contribution comes from the overlapping integral of $P\times S'$, which is $251$ {keV}, very close to the non-relativistic result $\Gamma_{non-rel}=254$ {keV}. The relativistic effect is estimated as
$$\frac{\Gamma_{rel}-\Gamma_{non-rel}}{\Gamma_{rel}}=6.62\%.$$
\begin{table}[H]
\begin{center}
\caption{Contributions of different partial waves to decay width (keV) of $\chi_{c2}(1P)$$\to$$\gamma\psi(1S)$.}\label{charm2++1-}
{\begin{tabular}{|c|c|c|c|c|} \hline\hline \diagbox {$2^{++}$}{$1^{--}$}
                          & ~whole~                      & ~$S'$ wave~                    & ~$P'$ wave~          & ~$D'$ wave~           \\ \hline
  ~whole~                   & 371                            & 269                           & 6.99                  & 0.0208             \\ \hline
  ~$P$ wave~                & 272                            & 260                          & 0.251            & ~$9.74\times10^{-4}$~    \\ \hline
  ~~~~$D$ wave~~~~      & 3.15                          & 0.548                          & 5.72        & ~~~~~$1.32\times10^{-4}$~~~~~\\ \hline
  ~$F$ wave~   & ~~~~~~$4.31\times10^{-3}$~~~~~~  & ~~~~~~$3.48\times10^{-6}$~~~~~~   & ~~~~~$3.69\times10^{-3}$~~~~~  & 0.0123           \\ \hline\hline
\end{tabular}}
\end{center}
\end{table}
Table \ref{charm2++1-} shows that the $\chi_{c2}(1P)$ is a $P$$-$$D$$-$$F$ mixing state, and $P$ wave provides the main contribution, $D$ wave has small contribution, $F$ wave has tiny contribution which can be ignored. The non-relativistic contribution is 293 keV,  a little larger than 260 keV from $P\times S'$, and the relativistic effect is $21.0\%.$

Our estimations of the relativistic effects indicate that the relativistic corrections are not very large in the decays of $\chi_{cJ}(1P)$$\to$$\gamma\psi(1S)$ $(J=0,1,2)$, so most of the theoretical decay widths from different models are comparable to each other, and some of them including ours agree well with experimental data.

\subsubsection{$\chi_{_{c0}}(2P) \to \psi+\gamma$}
$\chi_{_{c0}}(2P)$ has three main radiative decay channels, $\chi_{_{c0}}(2P) \to \psi(1S)+\gamma$, $\chi_{_{c0}}(2P) \to \psi(2S)+\gamma$ and $\chi_{_{c0}}(2P) \to \psi(3770)+\gamma$, where $\chi_{_{c0}}(2P) \to \psi(2S)+\gamma$ contributes the largest partial width, see Table \ref{charm}. We also notice that the results of different theoretical models vary greatly, for example, $\Gamma(\chi_{_{c0}}(2P) \to \psi(2S)+\gamma)$ varies from 44 keV to 135 keV. As mentioned earlier, in the case of highly excited states, relativistic corrections become significant. In Table \ref{charm2P0++1-}, we present the contributions of different partial waves for decay $\chi_{_{c0}}(2P) \to \psi(1S)+\gamma$. As can be seen, the decay width from $P\times S'$ is $15.8$ keV, which differs greatly from the whole result $45.1$ keV, indicating a large relativistic effect in this transition. The calculated non-relativistic decay width is $\Gamma_{non-rel}$=22.7 \rm{keV}, and the relativistic effect is $49.7\%$, which is much larger than the $23.5\%$ of $\chi_{_{c0}}(1P) \to \psi(1S)+\gamma$.

Another sensitive quantity that affects the results is the mass of the $\chi_{_{c0}}(2P)$. In Table \ref{charm}, different $\chi_{_{c0}}(2P)$ masses were used, for example, the mass of $\chi_{_{c0}}(2P)$ is 3852 MeV in the NR potential model \cite{Barnes_2005}, 3916 MeV in the relativistic GI model \cite{Barnes_2005}, 3842 MeV in screened potential model \cite{Li:2009zu} and 3862 MeV in our calculation, etc. Different mass of $\chi_{_{c0}}(2P)$ may has a greater impact on the result than relativistic correction.
Because there is a node in the $\chi_{_{c0}}(2P)$ (similar of $\psi(2S)$) wave function, the contributions of the wave functions on two sides of the node to the amplitude have opposite sign, and cancel to each other. The mass value of the $2P$ state affects the position of node, thus having a significant impact on the results. For example, in our calculation, the mass of $\chi_{_{c0}}(2P)$ is chosen as 3862 MeV, so we label this particle as $\chi_{_{c0}}(3860)$. If we take the mass of 3921.7 MeV, this is the mass of $\chi_{_{c0}}(3915)$, and its quantum number is also $0^{++}$. Then we get $\Gamma(\chi_{_{c0}}(3915) \to \psi(1S)+\gamma)=45.3$ keV, $\Gamma(\chi_{_{c0}}(3915) \to \psi(2S)+\gamma)=141$ keV, $\Gamma(\chi_{_{c0}}(3915) \to \psi(3770)+\gamma)=59.6$ keV. Except the first one, the later two are much different from the cases of $\chi_{_{c0}}(3860)\to \{\psi(2S),\psi(3770)\}$, which are $55.7$ keV and $12.8$ keV, respectively.
These values of decay widths, as well as their ratios,
$$\frac{\Gamma(\chi_{_{c0}}(3860) \to \psi(2S)+\gamma)}{\Gamma(\chi_{_{c0}}(3860) \to \psi(1S)+\gamma)}=1.24,~~
\frac{\Gamma(\chi_{_{c0}}(3860) \to \psi(1D)+\gamma)}{\Gamma(\chi_{_{c0}}(3860) \to \psi(1S)+\gamma)}=0.284,$$
$$\frac{\Gamma(\chi_{_{c0}}(3915) \to \psi(2S)+\gamma)}{\Gamma(\chi_{_{c0}}(3915) \to \psi(1S)+\gamma)}=3.11,~~\frac{\Gamma(\chi_{_{c0}}(3915) \to \psi(1D)+\gamma)}{\Gamma(\chi_{_{c0}}(3915) \to \psi(1S)+\gamma)}=1.32,$$
are crucial to determine the possibility of $\chi_{_{c0}}(3860)$ and $\chi_{_{c0}}(3915)$ being pure charmonia if we have experimental data.
\begin{table}[H]
\begin{center}
\caption{Contributions of different partial waves to decay width (keV) of $\chi_{c0}(3860)$$\to$$\gamma\psi(1S)$.}\label{charm2P0++1-}
{\begin{tabular}{|c|c|c|c|c|} \hline\hline \diagbox {$0^{++}$}{$1^{--}$}
                      & ~whole~    & ~$S'$ wave~   & ~$P'$ wave~          & ~$D'$ wave~             \\ \hline
  ~whole~              & 45.1         & 25.5 & ~~~~~~~~0.0133~~~~~~~~      & ~0.0387~               \\ \hline
~~~~~~~$P$ wave~~~~~~~ & 12.0  &~~~~~~~~15.8~~~~~~~~   & 2.09  &~~~~~~~~$4.86\times10^{-3}$~~~~~~~~ \\ \hline
  ~$S$ wave~    &~~~~~~~~10.6~~~~~~~~   &1.15          & 2.44              & ~0.0161~               \\ \hline\hline
  \end{tabular}}
\end{center}
\end{table}
\begin{table}[H]
\begin{center}
\caption{Contributions of different partial waves to decay width (keV) of $\chi_{c0}(3860)$$\to$$\gamma\psi(3770)$.}\label{charm2P1D0++1-}
{\begin{tabular}{|c|c|c|c|c|} \hline\hline \diagbox {$0^{++}$}{$1^{--}$}
                        & ~whole~                  & ~$D'$ wave~          & ~$P'$ wave~             & ~$S'$ wave~            \\ \hline
  ~whole~                & 12.8                       & 9.88               & ~0.129~          & ~$9.33\times10^{-4}$~        \\ \hline
~~~~~~~$P$ wave~~~~~~~   & 10.9                       & 9.77           &~~~~~~0.0140~~~~~~    & ~$3.43\times10^{-4}$~        \\ \hline
  ~$S$ wave~      & ~~~~~~0.0782~~~~~~  & ~~~~~~~$2.72\times10^{-4}$~~~~~~~ &~0.0580~    &~~~~~~~$1.45\times10^{-4}$~~~~~~~  \\ \hline\hline
\end{tabular}}
\end{center}
\end{table}

It is well known that $\psi(3770)$ is not a pure $D'$ wave, but a $S'-D'$ mixing state. While in our method, we find its wave function also has the component of $P'$ wave, and it is a $S'$$-$$P'$$-$$D'$ mixing state. Table \ref{charm2P1D0++1-} shows the partial wave contributions to the transition of $\chi_{c0}(3860)$$\to$$\gamma\psi(3770)$. We can see that $D'$ wave in $\psi(3770)$ provides the main contribution, $P'$ wave has a small contribution, while the contribution of $S'$ wave is very small and can be ignored. The non-relativistic contribution also comes mainly from $P$$\times$$D'$, and the relativistic effect is about $23.7\%$.

\subsubsection{$\chi_{_{c1}}(2P) \to \psi+\gamma$}
Similar to $\chi_{_{c0}}(2P)$, $\chi_{_{c1}}(2P)$ also has three radiative decay channels. We take the mass of $\chi_{_{c1}}(2P)$ as 3872 MeV, and its radiative decay widths are $\Gamma(\{\psi(1S),\psi(2S),\psi(1D)\}+\gamma)=\{39.2,69.4,4.44\}$ keV. Our result 39.2 keV of $\chi_{_{c1}}(3872) \to \psi(1S)+\gamma$ is much larger than the current experimental data $9.5\pm6.4$ keV, while the value 69.4 keV of $\chi_{_{c1}}(3872) \to \psi(2S)+\gamma$ is consistent with data $54\pm33$. Experiment also detected the following ratio \cite{PDG},
$$\frac{\Gamma(\chi_{_{c1}}(3872) \to \psi(2S)+\gamma)}{\Gamma(\chi_{_{c1}}(3872) \to \psi(1S)+\gamma)}=2.6\pm 0.6.$$
Our theoretical result
$$\frac{\Gamma(\chi_{_{c1}}(3872) \to \psi(2S)+\gamma)}{\Gamma(\chi_{_{c1}}(3872) \to \psi(1S)+\gamma)}=1.77$$ closes to the lower limit, so if $\chi_{_{c1}}(3872)$ is the conventional charmonium $\chi_{_{c1}}(2P)$ is still an open question.
We also obtain the ratio
$$\frac{\Gamma(\chi_{_{c1}}(3872) \to \psi(1D)+\gamma)}{\Gamma(\chi_{_{c1}}(3872) \to \psi(1S)+\gamma)}=0.113,$$ which may serve as an optional criterion for determining whether $\chi_{_{c1}}(3872)$ is a traditional charmonium.

Similarly, the decay width of $\chi_{_{c1}}(2P)$ is highly sensitive to its mass. If we take our  predicted theoretical mass of $3928.7$ MeV \cite{Chang:2010kj}, then the radiative decay widths become to $\Gamma(\{\psi(1S),\psi(2S),\psi(1D)\}+\gamma)=\{48.6,153,15.9\}$ keV, and the obtained ratios are
$$\frac{\Gamma(\chi_{_{c1}}(3929) \to \psi(2S)+\gamma)}{\Gamma(\chi_{_{c1}}(3929) \to \psi(1S)+\gamma)}=
3.15,~~\frac{\Gamma(\chi_{_{c1}}(3929) \to \psi(1D)+\gamma)}{\Gamma(\chi_{_{c1}}(3929) \to \psi(1S)+\gamma)}=0.327.$$
All these results differ significantly from those obtained with a mass of 3872 MeV. In other theoretical models, different $\chi_{_{c1}}(2P)$ masses are used, such as 3925 MeV in NR model \cite{Barnes_2005}, 3953 MeV in GI model \cite{Barnes_2005}, 3950 MeV in Ref.\cite{Radford:2007vd}, etc. Accordingly, we note that significant differences in the theoretical predictions of $\chi_{_{c1}}(2P)$ decays between different models.

\begin{table}[H]
\begin{center}
\caption{Contributions of different partial waves to decay width (keV) of $\chi_{c1}(3872)$$\to$$\gamma\psi(1S)$.}\label{charm2P1++1-}
{\begin{tabular}{|c|c|c|c|c|}\hline \hline \diagbox {$1^{++}$}{$1^{-}$}
                         & ~whole~   & ~$S'$ wave~    & ~$P'$ wave~         & ~$D'$ wave~             \\ \hline
  ~whole~                 & 39.2        & 14.9           & 2.43              & 0.0361                 \\ \hline
 ~~~~~~~~~$P$ wave~~~~~~~~~&15.8 &~~~~~~~~18.0~~~~~~~~   & 1.58 &~~~~~~~~$5.01\times10^{-3}$~~~~~~~~  \\ \hline
  ~$D$ wave~       &~~~~~~~~5.41~~~~~~~~ &0.445  & ~~~~~~~~6.95~~~~~~~~       &0.0166                 \\ \hline\hline
\end{tabular}}
\end{center}
\end{table}

\begin{table}[H]
\begin{center}
\caption{Contributions of different partial waves to  decay width (keV) of $\chi_{c1}(3872)$$\to$$\gamma\psi(2S)$.}\label{charm2P2S1++1-2}
{\begin{tabular}{|c|c|c|c|c|} \hline\hline \diagbox {$1^{++}$}{$1^{--}$}
                         & ~whole~                & ~$S'$ wave~           & ~$P'$ wave~        & ~$D'$ wave~             \\ \hline
  ~whole~                & 69.4                     & 61.0                & 0.183         & $5.19\times10^{-4}$          \\ \hline
 ~~~~~~~$P$ wave~~~~~~~  & 66.2                     & 61.3          & ~~~~~~0.0531~~~~~~   &$1.34\times10^{-4}$          \\ \hline
  ~$D$ wave~        &~~~~~~0.0361~~~~~~ &~~~~~~~$1.22\times10^{-3}~~~~~~~$ &0.0409  &~~~~~~~$1.36\times10^{-4}$~~~~~~~   \\ \hline\hline
\end{tabular}}
\end{center}
\end{table}
In Tables \ref{charm2P1++1-} and \ref{charm2P2S1++1-2}, we show the different partial wave contributions to the decay widths of $\chi_{c1}(3872)$ radiative decay to $\psi(1S)$ and $\psi(2S)$, respectively. Table \ref{charm2P1++1-} shows that each independent contribution is much smaller than the whole width, which means that the cross term contributions that are not listed are also significant, indicating a large relativistic correction in this process. The non-relativistic contribution is $\Gamma_{non-rel}$=27.1 \rm{keV} in $\chi_{c1}(3872)$$\to$$\gamma\psi(1S)$, and the relativistic effect is
$30.9\%$. While in Table \ref{charm2P2S1++1-2}, the $P\times S'$ provides the dominant contribution to $\chi_{c1}(3872)$$\to$$\gamma\psi(2S)$, the corresponding non-relativistic result is $\Gamma_{non-rel}$=63.1 \rm{keV}, and the relativistic effect is $9.08\%$.

\subsubsection{$\chi_{_{c2}}(2P) \to \psi+\gamma$}
Unlike $\chi_{c0}(2P)$ and $\chi_{c1}(2P)$, the candidate for $\chi_{c2}(2P)$ is only $\chi_{c2}(3930)$ and there is not much controversy. Its mass is 3922.5 MeV.
From Tables \ref{charm2P2++1-} and \ref{charm2P2S2++1-}, it can be seen that in our method, $\chi_{c2}$ is a $P$ wave dominant $P$$-$$D$$-$$F$ mixture. $P$ wave provides the main contribution, $D$ wave give small but sizable contribution, while the contribution of $F$ wave is very small. The relativistic effects for $\chi_{c2}(2P)$$\to$$\gamma\psi(1S)$ and $\chi_{c2}(2P)$$\to$$\gamma\psi(2S)$ are
$37.5\%$ and $19.3\%$, respectively.

\begin{table}[H]
\begin{center}
\caption{Contributions of different partial waves to decay width (keV) of $\chi_{c2}(2P)$$\to$$\gamma\psi(1S)$.}\label{charm2P2++1-}
{\begin{tabular}{|c|c|c|c|c|} \hline\hline \diagbox {$2^{++}$}{$1^{--}$}
                        & ~whole~           & ~$S'$ wave~          & ~$P'$ wave~       & ~$D'$ wave~           \\ \hline
         ~whole~          & 59.5               & 13.0         & ~~~~~~~9.10~~~~~~~      & 0.0567               \\ \hline
~~~~~~~~~$P$ wave~~~~~~~~~& 17.1               & 13.9               & 0.351  &~~~~~~~$1.86\times10^{-3}$~~~~~~~\\ \hline
       ~$D$ wave~  & ~~~~~~~4.54~~~~~~~        & 3.28                & 12.8       & ~$1.84\times10^{-3}$~      \\ \hline
       ~$F$ wave~        & 0.0182 & ~~~~~~~$1.34\times10^{-4}$~~~~~~~&0.0499            & 0.0640               \\ \hline\hline
\end{tabular}}
\end{center}
\end{table}

\begin{table}[H]
\begin{center}
\caption{Contributions of different partial waves to decay width (keV) of $\chi_{c2}(2P)$$\to$$\gamma\psi(2S)$.}\label{charm2P2S2++1-}
{\begin{tabular}{|c|c|c|c|c|} \hline\hline \diagbox {$2^{++}$}{$1^{-}$}
                             & ~whole~                     & ~$S'$ wave~           & ~$P'$ wave~            & ~$D'$ wave~            \\ \hline
         ~whole~             & 140                         & 108                      & 1.56              & $3.64\times10^{-3}$~     \\ \hline
 ~~~~~~~$P$ wave~~~~~~~      & 107                         & 102                    & 0.0565          &~~~~~$2.83\times10^{-5}$~~~~~ \\ \hline
       ~$D$ wave~            & 1.01                      & 0.0132                     & 1.21              & $1.60\times10^{-4}$~     \\ \hline
       ~$F$ wave~&~~~~~$3.17\times10^{-3}$~~~~~&~~~~~$3.85\times10^{-7}$~~~~~&~~~~~$3.73\times10^{-5}$~~~~~&$3.82\times10^{-3}$~     \\ \hline\hline
\end{tabular}}
\end{center}
\end{table}
\subsubsection{$\chi_{_{c2}}(1F)\rightarrow\gamma\psi$}

Similar to $\chi_{c2}(1P)$ and $\chi_{c2}(2P)$, $\chi_{c2}(1F)$ is also a $P$$-$$D$$-$$F$ mixing state, but the difference is that it is $F$-wave dominant state. Its radiative decay modes are the same as $\chi_{c2}(2P)$, and also can decay to $\psi(1S)$, $\psi(1S)$, or $\psi(3770)$. Among them, the partial width of decay $\chi_{c2}(1F)$$\to$$\gamma\psi(3770)$ is the largest. We show the contributions of different partial waves to decay width of this process in Table \ref{charm1F1D2++1-}, where we can see that $F$ wave in $\chi_{c2}(1F)$ is the dominant partial wave, and contributes the most to the decay width, $D$ wave has a small contribution, and the $P$ wave can be ignored. From Table \ref{charm1F1D2++1-}, we once again note that $\psi(3770)$ is a $D$ wave dominated $S-P-D$ mixing state, where the contribution of $P$ wave is small and the $S$ wave can be ignored. The non-relativistic decay width is 285 keV from $F\times D'$, and the relativistic effect is $15.7\%$.

Currently, $\chi_{c2}(1F)$ is not discovered by experiment,
in our calculation, the used mass of $\chi_{c2}(1F)$ is 4027 MeV, which is predicted by our model, if we change it to 4090 MeV, the obtained decay widths of $\chi_{c2}(1F)$$\to$$\{\psi(1S),~\psi(2S),~\psi(3770)\}+\gamma$ are changed from $\{4.30,~0.242,~338\}$ keV to $\{5.34,~0.433,~561\}$ keV. Compared with cases of $\chi_{c0}(2P)$ and $\chi_{c1}(2P)$, the changes are not much, this may due to the wave function of $\chi_{c2}(1F)$ does not contain node structure, except the channel of $\chi_{c2}(1F)\to\psi(2S)+\gamma$, which has a node in final $\psi(2S)$, the other two is affected by the phase space, not from node structure, so the decay widths are not very sensitive to the mass value of $\chi_{c2}(1F)$. The channel decaying to $\psi(3770)$ changes a little more, because some of its form factors are multiply by two $P_f$ ($P_f$ eventually becomes $\vec P_f$), see the last formula in Eq.(\ref{shape}), so it is more sensitive to the phase space.
\begin{table}[H]
\begin{center}
\caption{Contributions of different partial waves to  decay width (keV) of $\chi_{c2}(1F)$$\to$$\gamma\psi(3770)$.}\label{charm1F1D2++1-}
{\begin{tabular}{|c|c|c|c|c|} \hline\hline \diagbox {$2^{++}$}{$1^{--}$}
                     & ~whole~                     & ~$D'$ wave~               & ~$P'$ wave~                       & ~$S'$ wave~               \\ \hline
      ~whole~          & 338                           & 288                       & 1.97                      & ~$1.82\times10^{-4}$~          \\ \hline
    ~$F$ wave~         & 282                           & 285              & ~~~~~~$4.59\times10^{-3}$~~~~~~     & ~$3.85\times10^{-4}$~          \\ \hline
~~~$D$ wave~~~       & 2.49                         & 0.0571                     & 2.14                 & ~~~~~~$1.38\times10^{-4}$~~~~~~     \\ \hline
    ~$P$ wave~&~~~~~$1.25\times10^{-3}$~~~~~&~~~~~~$1.25\times10^{-4}$~~~~~~   & ~$5.99\times10^{-5}$~          & ~$2.35\times10^{-4}$~          \\ \hline\hline
\end{tabular}}
\end{center}
\end{table}
\subsubsection{$\chi_{_{cJ}}(3P) \to \psi+\gamma$ $(J=0,1,2)$}
At present, no highly excited states $\chi_{_{cJ}}(3P)(J=0,1,2)$ have been found in experiments. Due to large masses, they have more decay channels. In most theoretical models, their main electromagnetic decay channels are decays to $\psi(3S)$. Since the wave functions of initial and final mesons both contain two node structures, the decay widths of $\chi_{_{cJ}}(3P) \to \psi(3S)+\gamma$ $(J=0,1,2)$ are sensitive to the masses of $\chi_{_{cJ}}(3P)(J=0,1,2)$. For example, compared with other models, we obtained the smallest decay width for $\chi_{_{c0}}(3P) \to \psi(3S)+\gamma$, 19.4 keV. The reason is that our predicting mass of $\chi_{_{c0}}(3P)$ is the smallest one, 4027 MeV. If we change its mass to 4200 MeV or 4250 MeV, then the width becomes to 80.9 keV or 186 keV accordingly, showing that the decay width is very sensitive to the mass or the phase space.

\subsection{The decays of $\chi_{_{bJ}}\to \Upsilon+\gamma$}
Currently, experiments have detected the masses of the $\chi_{_{bJ}}(1P)$ and $\chi_{_{bJ}}(2P)$, where $(J=0,1,2)$. And the branching ratios of $\chi_{_{bJ}}(1P)\to \Upsilon(1S)+\gamma$ and $\chi_{_{bJ}}(2P)\to \Upsilon(1S)+\gamma$ as well as $\chi_{_{bJ}}(2P)\to \Upsilon(2S)+\gamma$ are also detected, but with large uncertainties. While the total decay widths of all $\chi_{_{bJ}}$ are still missing, we have no partial decay widths from experiments.
For highly excited states, including $\chi_{_{bJ}}(3P)$, $\chi_{_{b2}}(1F)$, $\Upsilon(1D)$ and $\Upsilon(2D)$,  they have not been confirmed or discovered by experiment.
Fortunately, due to the relatively small relativistic corrections of bottomonia, the credibility of theoretical predictions has greatly increased, including some results from non-relativistic models.

\begin{table}[htp]
\begin{center}
\caption{The decay widths of $\chi_{_{bJ}}\to\gamma\Upsilon$ in unit of keV.}\label{bottom}
\begin{tabular}{c c c c c c c  cc c}
\hline\hline
\textbf{~~~Process~~~} &\textbf{~~~Ours~~~}&\cite{Ebert:2002pp}&~~~\cite{Radford:2007vd}~~~ &~~~\cite{Godfrey:2015dia}~~~ &~~~\cite{Segovia:2016xqb}~~~&~~~\cite{Barducci:2016wze}~~~ &~~~\cite{Deng:2016ktl}~~~ &~~~\cite{Ananyev:2020uve}~~~       \\
\hline
$\chi_{_{b0}}(1P)\rightarrow\gamma\Upsilon(1S)$ &28.3    &29.9 &19.6 &23.8 &28.07  &19.65 &27.5  &24.2 \\
$\chi_{_{b1}}(1P)\rightarrow\gamma\Upsilon(1S)$ &34.9    &36.6 &23.9 &29.5 &35.66  &29.48 &31.9  &30.2 \\
$\chi_{_{b2}}(1P)\rightarrow\gamma\Upsilon(1S)$ &39.9    &40.2 &26.3 &32.8 &39.15  &23.73 &31.8  &36.1 \\
\hline
$\chi_{_{b0}}(2P)\rightarrow\gamma\Upsilon(1S)$ &5.83    &6.79 &1.83 &2.5  &5.44   &1.49  &5.54  &4.4  \\
$\chi_{_{b0}}(2P)\rightarrow\gamma\Upsilon(2S)$ &11.9    &11.0 &9.91 &10.9 &12.80  &11.77 &14.4  &13.2 \\
$\chi_{_{b0}}(2P)\rightarrow\gamma\Upsilon(1D)$ &2.39    &1.17 &1.05 &1.0  &0.74   &      &1.77  &     \\
\hline
$\chi_{_{b1}}(2P)\rightarrow\gamma\Upsilon(1S)$ &6.64    &7.49 &4.81 &5.5  &9.13   &7.68  &10.8  &10.7 \\
$\chi_{_{b1}}(2P)\rightarrow\gamma\Upsilon(2S)$ &15.6    &14.7 &12.4 &13.3 &15.89  &46.80 &15.3  &15.3 \\
$\chi_{_{b1}}(2P)\rightarrow\gamma\Upsilon(1D)$ &0.998   &0.615&0.52 &0.5  &0.41   &      &0.56  &     \\
\hline
$\chi_{_{b2}}(2P)\rightarrow\gamma\Upsilon(1S)$ &7.38    &8.02 &6.86 &8.4  &11.38  &10.27 &12.5  &16.3 \\
$\chi_{_{b2}}(2P)\rightarrow\gamma\Upsilon(2S)$ &17.8    &16.7 &13.5 &14.3 &17.50  &18.77 &15.3  &16.7 \\
$\chi_{_{b2}}(2P)\rightarrow\gamma\Upsilon(1D)$ &0.0555  &0.035&0.03 &0.03 &0.0209 &      &0.026 &     \\
\hline
$\chi_{_{b2}}(1F)\rightarrow\gamma\Upsilon(1S)$ &0.0607  &     &     &     &       &       &     &     \\
$\chi_{_{b2}}(1F)\rightarrow\gamma\Upsilon(2S)$ &0.00345 &     &     &     &       &       &     &     \\
$\chi_{_{b2}}(1F)\rightarrow\gamma\Upsilon(1D)$ &34.6    &     &     &16.4 &       &       &     &     \\
$\chi_{_{b2}}(1F)\rightarrow\gamma\Upsilon(3S)$ &0.000457&     &     &     &       &       &     &     \\
\hline
$\chi_{_{b0}}(3P)\rightarrow\gamma\Upsilon(1S)$ &2.19    &     &     &0.3  &1.99   &      &1.87  &1.4  \\
$\chi_{_{b0}}(3P)\rightarrow\gamma\Upsilon(2S)$ &3.98    &     &     &1.7  &2.99   &      &2.55  &2.2  \\
$\chi_{_{b0}}(3P)\rightarrow\gamma\Upsilon(1D)$ &0.0631  &     &     &0.20 &0.0359 &      &0.15  &     \\
$\chi_{_{b0}}(3P)\rightarrow\gamma\Upsilon(3S)$ &11.4    &     &     &6.9  &8.50   &      &7.95  &7.6  \\
$\chi_{_{b0}}(3P)\rightarrow\gamma\Upsilon(2D)$ &3.99    &     &     &1.0  &3.50   &      &2.20  &     \\
\hline
$\chi_{_{b1}}(3P)\rightarrow\gamma\Upsilon(1S)$ &2.48    &     &     &1.3  &4.17   &      &6.41  &5.5  \\
$\chi_{_{b1}}(3P)\rightarrow\gamma\Upsilon(2S)$ &4.71    &     &     &3.1  &4.58   &      &5.63  &5.0  \\
$\chi_{_{b1}}(3P)\rightarrow\gamma\Upsilon(1D)$ &0.0429  &     &     &0.007&0.0480 &      &0.010 &     \\
$\chi_{_{b1}}(3P)\rightarrow\gamma\Upsilon(3S)$ &19.2    &     &     &8.4  &9.62   &      &10.3  &9.4  \\
$\chi_{_{b1}}(3P)\rightarrow\gamma\Upsilon(2D)$ &2.03    &     &     &0.47 &1.26   &      &1.07  &     \\
\hline
$\chi_{_{b2}}(3P)\rightarrow\gamma\Upsilon(1S)$ &2.79    &     &     &2.8  &5.65   &       &8.17 &10.7 \\
$\chi_{_{b2}}(3P)\rightarrow\gamma\Upsilon(2S)$ &4.95    &     &     &4.5  &5.62   &       &6.72 &7.5  \\
$\chi_{_{b2}}(3P)\rightarrow\gamma\Upsilon(1D)$ &0.00513 &     &     &     &0.00338&       &0.047&     \\
$\chi_{_{b2}}(3P)\rightarrow\gamma\Upsilon(3S)$ &20.5    &     &     &9.3  &10.38  &       &10.8 &11.2 \\
$\chi_{_{b2}}(3P)\rightarrow\gamma\Upsilon(2D)$ &0.126   &     &     &0.027&0.18   &       &0.049&     \\
\hline\hline
\end{tabular}
\end{center}
\end{table}

Our results on the electromagnetic decay widths of bottomonia $\chi_{_{bJ}}$ are presented in Table \ref{bottom}, where we also show some other theoretical results for comparison. Compared to the cases of charmonia, most theoretical results of the $\chi_{_{bJ}}(1P)$ and $\chi_{_{bJ}}(2P)$ decays agree well with each other. This is because the mass of the bottomonium is much heavier than that of the charmonium, so its relativistic correction is very small. Meanwhile, all the masses of $\chi_{_{bJ}}(1P)$ and $\chi_{_{bJ}}(2P)$ have been detected by experiments, there is no uncertainty caused by mass value. Among the low excited state, only the $\chi_{_{bJ}}(1D)$ are still missing, so the decays including $\chi_{_{bJ}}(1D)$ do not fit very well among different theories. Similarly, for highly excited states that have not been detected, such as $\chi_{_{bJ}}(3P)$ and $\chi_{_{b2}}(1F)$, due to the different masses in different models, the theoretical predictions of decay widths don't match very well.

\subsubsection{$\chi_{_{bJ}}(1P)\rightarrow\gamma\Upsilon(1S)$}
We pointed out above that the good agreement between decay widths of $\chi_{_{bJ}}(1P)\rightarrow\gamma\Upsilon(1S)$ $(J=0,1,2)$ predicted by different models are due to the small relativistic corrections in these transitions. To see these, we present Tables \ref{bottom0++1-}, \ref{bottom1++1-}, \ref{bottom2++1-}, in which we show the contributions of different partial waves to the decay widths. Among these Tables, we can see that the biggest contributions come from the $P\times S'$, indicating that non-relativistic results give the dominant contributions. We also found that due to the small relativistic correction, the contribution of the $P\times S'$ is almost equal to the non-relativistic result. In Table \ref{bottom0++1-}, the contribution of $P\times S'$ is 25.8 keV, the non-relativistic result we calculated is 25.9 keV, and the relativistic effect is
$8.48\%$ for $\chi_{_{b0}}(1P)\rightarrow\gamma\Upsilon(1S)$. Similarly, we obtain $\Gamma_{non-rel}$=34.1 keV ($=P\times S'$) and a very small relativistic effect
$2.29\%$ for $\chi_{_{b1}}(1P)\rightarrow\gamma\Upsilon(1S)$. $\Gamma_{non-rel}$=37.4 keV and relativistic effect
$6.27\%$ for $\chi_{_{b2}}(1P)\rightarrow\gamma\Upsilon(1S)$.

\begin{table}[htp]
\begin{center}
\caption{Contributions of different partial waves to decay width (\rm{keV}) of $\chi_{_{b0}}(1P)$$\to $$\gamma\Upsilon(1S)$.}\label{bottom0++1-}
{\begin{tabular}{|c|c|c|c|c|}
\hline\hline
\diagbox{$0^{++}$}{$1^{-}$}
                         &~whole~     & ~$S'$ wave~                & ~$P'$ wave~              & ~$D'$ wave~             \\ \hline
  ~whole~                & 28.3          & 29.3                      & 0.0101             & ~$1.45\times 10^{-5}$~      \\ \hline
  ~~~~~~~$P$ wave~~~~~~~ & 25.8   & ~~~~~~~25.8~~~~~~~ & ~~~~~~~$1.45\times 10^{-8}$~~~~~~~&~$1.11\times 10^{-5}$~      \\ \hline
  ~$S$ wave~      & ~~~~~~~0.0553~~~~~~~ & 0.106                     & 0.0100       & ~~~~~~~$2.24\times 10^{-7}$~~~~~~~\\ \hline\hline
\end{tabular}}
\end{center}
\end{table}
\begin{table}[H]
\begin{center}
\caption{Contributions of different partial waves to decay width (\rm{keV}) of $\chi_{_{b1}}(1P)$$\to$$ \gamma\Upsilon(1S)$.}\label{bottom1++1-}
{\begin{tabular}{|c|c|c|c|c|}
\hline\hline
\diagbox{$1^{++}$}{$1^{-}$}&~whole~                   & ~$S'$ wave~                     & ~$P'$ wave~              & ~$D'$ wave~         \\ \hline
  ~whole~                 & 34.9                          & 33.8                           & 0.103           & ~$3.24\times 10^{-6}$~    \\ \hline
~~~~~~$P$ wave~~~~~~      & 35.2                          & 34.1                           & 0.105       & ~~~~~$7.93\times 10^{-6}$~~~~~\\ \hline
  ~$D$ wave~ & ~~~~~$1.25\times 10^{-3}$~~~~~& ~~~~~$1.33\times 10^{-3}$~~~~~ &  ~~~~~$1.40\times 10^{-6}$~~~~~&$1.38\times 10^{-6}$~    \\ \hline\hline
\end{tabular}}
\end{center}
\end{table}
\begin{table}[H]
\begin{center}
\caption{Contributions of different partial waves to decay width (keV) of $\chi_{_{b2}}(1P)$$\to$$ \gamma\Upsilon(1S)$.}\label{bottom2++1-}
{\begin{tabular}{|c|c|c|c|c|}
\hline\hline
\diagbox{$2^{++}$}{$1^{-}$}
                        &~whole~               & ~$S'$ wave~                 & ~$P'$ wave~                & ~$D'$ wave~        \\ \hline
  ~whole~               & 39.9                    & 36.4                     & 0.0820             & ~$2.69\times 10^{-5}$~     \\ \hline
  ~$P$ wave~            & 36.7                    & 36.2               & ~$4.01\times 10^{-3}$~   & ~$1.80\times 10^{-6}$~     \\ \hline
~~~~~~~$D$ wave~~~~~~~ & 0.0279       & ~~~~~$8.82\times 10^{-3}$~~~~~       & 0.0608         & ~~~~~$2.15\times 10^{-6}$~~~~~ \\ \hline
  $F$ wave&   ~~~~~$4.29\times 10^{-6}$~~~~~&$9.01\times 10^{-10}$~& ~~~~~$4.72\times 10^{-6}$~~~~~&~$1.33\times 10^{-5}$~     \\ \hline\hline
\end{tabular}}
\end{center}
\end{table}

\subsubsection{$\chi_{_{bJ}}(2P)\rightarrow\gamma\Upsilon$}
$\chi_{_{bJ}}(2P)$ $(J=0,1,2)$ has three electromagnetic decay channels, namely $\chi_{_{bJ}}(2P)\rightarrow\gamma\Upsilon(1S)$, $\chi_{_{bJ}}(2P)\rightarrow\gamma\Upsilon(2S)$, and $\chi_{_{bJ}}(2P)\rightarrow\gamma\Upsilon(1D)$. Among them, $\chi_{_{bJ}}(2P)\rightarrow\gamma\Upsilon(2S)$ gives the maximum branching ratio.
In the experiment, the decay widths of these processes are not available. Instead, there are some branching ratios in experiment, from which we can obtain the following ratios \cite{PDG},
$$\frac{\Gamma_{ex}(\chi_{_{b0}}(2P)\to\gamma\Upsilon(2S))}{\Gamma_{ex}(\chi_{_{b0}}(2P)\to\gamma\Upsilon(1S))}=3.6^{+4.4}_{-1.6},$$ $$\frac{\Gamma_{ex}(\chi_{_{b1}}(2P)\to\gamma\Upsilon(2S))}{\Gamma_{ex}(\chi_{_{b1}}(2P)\to\gamma\Upsilon(1S))}=1.8^{+0.4}_{-0.3},$$ $$\frac{\Gamma_{ex}(\chi_{_{b2}}(2P)\to\gamma\Upsilon(2S))}{\Gamma_{ex}(\chi_{_{b2}}(2P)\to\gamma\Upsilon(1S))}=1.3^{+0.4}_{-0.3}.$$
Our predictions are, $$\frac{\Gamma(\chi_{_{b0}}(2P)\to\gamma\Upsilon(2S))}{\Gamma(\chi_{_{b0}}(2P)\to\gamma\Upsilon(1S))}=2.04,$$ $$\frac{\Gamma(\chi_{_{b1}}(2P)\to\gamma\Upsilon(2S))}{\Gamma(\chi_{_{b1}}(2P)\to\gamma\Upsilon(1S))}=2.34,$$ $$\frac{\Gamma(\chi_{_{b2}}(2P)\to\gamma\Upsilon(2S))}{\Gamma(\chi_{_{b2}}(2P)\to\gamma\Upsilon(1S))}=2.41.$$ The first two are within the experimental errors, and the last one is slightly larger than the experimental data.

\begin{table}[H]
\begin{center}
\caption{Contributions of different partial waves to decay width (\rm{keV}) of $\chi_{_{b0}}(2P)$$\to$$ \gamma\Upsilon(1S)$.}\label{bottom2P0++1-}
{\begin{tabular}{|c|c|c|c|c|}
\hline\hline
\diagbox{$0^{++}$}{$1^{-}$}
                       &~whole~   & ~$S'$ wave~        & ~$P'$ wave~               & ~$D'$ wave~              \\ \hline
  ~whole~              & 5.83        & 5.47 &~~~~~~~$1.61\times 10^{-3}$~~~~~~~& ~$5.36\times 10^{-5}$~       \\ \hline
~~~~~~~~$P$ wave~~~~~~~~&3.95 & ~~~~~~~4.42~~~~~~~      & 0.0410          &~~~~~~~$4.33\times 10^{-7}$~~~~~~~ \\ \hline
  ~$S$ wave~    &~~~~~~~0.184~~~~~~~ &0.0556            & 0.0264               & ~$6.36\times 10^{-5}$~       \\ \hline\hline
\end{tabular}}
\end{center}
\end{table}

In the cases of charmonia, we found that processes $\chi_{_{cJ}}(2P)\to\gamma\psi(1S)$ have large relativistic effects. Here, as comparisons, we are willing to study the partial wave contributions and relativistic effects of the $\chi_{_{bJ}}(2P)\rightarrow\gamma\Upsilon(1S)$ $(J=0,1,2)$. In Tables \ref{bottom2P0++1-}, \ref{bottom2P1++1-} and \ref{bottom2P2++1-}, we show the results of different partial waves contributing to decay widths. It can be seen that the main contributions also come from $P\times S'$, whose values are close to the non-relativistic results. We also note that the $D$ wave in $\chi_{_{b1}}(2P)$ and $\chi_{_{b2}}(2P)$, $S$ wave in $\chi_{_{b0}}(2P)$, as well as $P'$ wave in $\Upsilon(1S)$ have much larger contributions in the decays of $\chi_{_{bJ}}(2P)\to\gamma\Upsilon(1S)$ than those of $\chi_{_{bJ}}(1P)\to\gamma\Upsilon(1S)$, which means the relativistic corrections are much larger in the former than in the later. For $\chi_{_{bJ}}(2P)\rightarrow\gamma\Upsilon(1S)$ $(J=0,1,2)$, we obtain the relativistic effects are $\{17.8\%,~7.08\%,~12.9\%\}$, which are indeed much larger than those of $\chi_{_{bJ}}(1P)\rightarrow\gamma\Upsilon(1S)$, $\{8.48\%,~2.29\%,~6.27\%\}$, but much smaller than the $\{49.7\%,~30.9\%,~37.5\%\}$ in the decays of $\chi_{_{cJ}}(2P)\to\gamma\psi(1S)$.

\begin{table}[H]
\begin{center}
\caption{Contributions of different partial waves to decay width (\rm{keV}) of $\chi_{_{b1}}(2P)$$\to$$ \gamma\Upsilon(1S)$.}\label{bottom2P1++1-}
{\begin{tabular}{|c|c|c|c|c|}
\hline\hline
\diagbox{$1^{++}$}{$1^{-}$}
                      &~whole~          & ~$S'$ wave~   & ~$P'$ wave~        & ~$D'$ wave~               \\ \hline
  ~whole~            & 6.64                & 5.38   &~~~~~~~0.0284~~~~~~~~  &$5.70\times 10^{-5}$        \\ \hline
~~~~~~$P$ wave~~~~~~ & 5.63                & 5.71          &0.0130  &~~~~~~~~$4.52\times 10^{-5}$~~~~~~~~\\ \hline
  ~$D$ wave  &~~~~~~~0.0464~~~~~~~ &~~~~~~~~0.0108~~~~~~~~ &0.0727          &$5.13\times 10^{-5}$        \\ \hline\hline
\end{tabular}}
\end{center}
\end{table}

\begin{table}[H]
\begin{center}
\caption{Contributions of different partial waves to decay width (\rm{keV}) of $\chi_{_{b2}}(2P)$$\to$$ \gamma\Upsilon(1S)$.}\label{bottom2P2++1-}
{\begin{tabular}{|c|c|c|c|c|}
\hline\hline
\diagbox{$2^{++}$}{$1^{-}$} &~whole~                     & ~$S'$ wave~                & ~$P'$ wave~            & ~$D'$ wave~              \\ \hline
  ~whole~                   & 7.38                          & 4.92                    & ~0.158~             & ~$1.08\times 10^{-4}$~      \\ \hline
  ~$P$ wave~                & 5.38                          & 5.13       & ~~~~~$3.11\times 10^{-3}$~~~~~    & $2.44\times 10^{-6}$~      \\ \hline
~~~~~~~$D$ wave~~~~~~~      &0.0567                         &0.0633                    &~0.192~         & ~~~~~$9.92\times 10^{-6}$~~~~~  \\ \hline
  ~$F$ wave~   & ~~~~~$2.59\times 10^{-5}$~~~~~&~~~~~$4.86\times 10^{-8}$~~~~~& $8.80\times 10^{-5}$        & ~$9.50\times 10^{-5}$~      \\ \hline\hline
\end{tabular}}
\end{center}
\end{table}

$\Upsilon(1D)$ has not been discovered, the decays of $\chi_{_{bJ}}(2P)\rightarrow\gamma\Upsilon(1D)$ $(J=0,1,2)$ can serve as candidate for discovery channels. Their decay widths are $\{2.39,~0.998,~0.0555\}$ keV. And we get the following ratios,
$$\frac{\Gamma(\chi_{_{b0}}(2P)\to\gamma\Upsilon(1D))}{\Gamma(\chi_{_{b0}}(2P)\to\gamma\Upsilon(1S))}=0.410,$$ $$\frac{\Gamma(\chi_{_{b1}}(2P)\to\gamma\Upsilon(1D))}{\Gamma(\chi_{_{b1}}(2P)\to\gamma\Upsilon(1S))}=0.150,$$ $$\frac{\Gamma(\chi_{_{b2}}(2P)\to\gamma\Upsilon(1D))}{\Gamma(\chi_{_{b2}}(2P)\to\gamma\Upsilon(1S))}=0.00752.$$
$\Upsilon(1D)$ has a sizable production rate in the decay of $\chi_{_{b0}}(2P)$. In experiments, $\Upsilon(1D)$ can be also directly generated in $B$ Factory through $e^+ e^-$ annihilation, it is possible to search for $\Upsilon(1D)$ through decay chain $\Upsilon(1D)\to \gamma\chi_{_{b0}}(1P)$, $\chi_{_{b0}}(1P)\rightarrow\gamma\Upsilon(1S)$ and $\Upsilon(1S)\to \mu^+ \mu^-$, or $\Upsilon(1D)\to \mu^+ \mu^-$.

In Table \ref{bottom2P1D0++1-}, we take $\chi_{_{b0}}(2P)$$\to$$\gamma\Upsilon(1D)$ as an example and provide the contributions of different partial waves to the decay width. It can be seen that, similar to $\psi(1D)$, $\Upsilon(1D)$ is also a $S'$$-$$P'$$-$$D'$ mixing state. The non-relativistic $D'$ wave provides the dominant contribution, the relativistic $P'$ wave has small contribution, and the tiny contribution of $S'$ wave is ignorable.
Table \ref{bottom2P1D0++1-} shows that the relativistic correction is small and the relativistic effect is calculated as $7.11\%$.

\begin{table}[H]
\begin{center}
\caption{Contributions of different partial waves to decay width (\rm{keV}) of $\chi_{_{b0}}(2P)$$\to$$ \gamma\Upsilon(1D)$.}\label{bottom2P1D0++1-}
{\begin{tabular}{|c|c|c|c|c|} \hline\hline \diagbox {$0^{++}$}{$1^{-}$}
                         & ~whole~                   & ~$D'$ wave~           & ~$P'$ wave~              & ~$S'$ wave~          \\ \hline
  ~whole~                & 2.39                        & 2.24           & ~$1.93\times10^{-4}$~     & ~$1.40\times10^{-5}$~    \\ \hline
~~~~~~~$P$ wave~~~~~~~   & 2.30                        & 2.22       & ~~~~~$3.21\times10^{-4}$~~~~~ & ~$1.04\times10^{-5}$~    \\ \hline
  ~$S$ wave~  &~~~~~$9.99\times10^{-4}$~~~~~&~~~~~$2.16\times10^{-5}$~~~~~&$6.77\times10^{-4}$~ & ~~~~~$2.57\times10^{-7}$~~~~~\\ \hline\hline
\end{tabular}}
\end{center}
\end{table}

\subsubsection{$\chi_{_{b2}}(1F)\rightarrow\gamma\Upsilon$}
Similar to $\chi_{_{c2}}(1F)$, $\chi_{_{b2}}(1F)$ is also a $P$$-$$D$$-$$F$ mixing state. Its wave function is dominated by $F$ wave, and mixed with $P$ and $D$ waves. Its electromagnetic decay channels include $\chi_{_{b2}}(1F)\rightarrow\gamma\Upsilon(nS)$ ($n=1,2,3$) and $\chi_{_{b2}}(1F)\rightarrow\gamma\Upsilon(1D)$, with the last one having the largest decay width, far greater than those of others.
Our prediction is 34.6 keV, almost twice the result of 16.4 keV in Ref.\cite{Godfrey:2015dia}, which is also a relativistic model. The difference between ours and theirs mainly comes from the phase space difference.
If we adjust the masses of $\chi_{_{b2}}(1F)$ and $\Upsilon(1D)$ from 10374 \rm{MeV} and 10129 \rm{MeV} (ours) to the values in Ref.\cite{Godfrey:2015dia}, which are 10350 \rm{MeV} and 10138 \rm{MeV}, then our calculated decay width becomes 22.6 \rm{keV}.
Both Ref.\cite{Godfrey:2015dia} and we predict that the mass of $\chi_{_{b2}}(1F)$ is below the $B\bar B$ threshold, therefore $\chi_{_{b2}}(1F)$ has no strong decays allowed by OZI rule, and EM decays are important channels for it. We believe that the decay chains $\chi_{_{b2}}(1F)\to \gamma\Upsilon(1D)$, $\Upsilon(1D)\to \gamma\chi_{_{b0}}(1P)$, $\chi_{_{b0}}(1P)\rightarrow\gamma\Upsilon(1S)$ and $\Upsilon(1S)\to \mu^+ \mu^-$, and $\chi_{_{b2}}(1F)\to \gamma\Upsilon(1D)$ followed by $\Upsilon(1D)\to \mu^+ \mu^-$ will play important roles in searching for $\chi_{_{b2}}(1F)$ in experiment.

In Table \ref{bottom1F1D2++1-}, we show the contributions of different partial waves to the decay width of $\chi_{_{b2}}(1F)\rightarrow\gamma\Upsilon(1D)$. It can be seen that,
the non-relativistic term $F\times D'$ has the dominant contribution and the relativistic effect is $5.20\%$.

\begin{table}[H]
\begin{center}
\caption{Contributions of different partial waves to decay width (\rm{keV}) of $\chi_{_{b2}}(1F)$$\to$$\gamma \Upsilon(1D)$.}\label{bottom1F1D2++1-}
{\begin{tabular}{|c|c|c|c|c|}
\hline\hline
\diagbox{$2^{++}$}{$1^{-}$}&~whole~                & ~$D'$ wave~               & ~$P'$ wave~                & ~$S'$ wave~           \\ \hline
  ~whole~                   & 34.6                    & 32.9                     & 0.0148              & ~$6.08\times 10^{-4}$~     \\ \hline
  ~$F$ wave~                & 33.0                    & 32.8                & ~$6.39\times 10^{-5}$~   & ~$5.74\times 10^{-4}$~     \\ \hline
~~~~~~~$D$ wave~~~~~~~     & 0.0198        & ~~~~~$7.02\times 10^{-4}$~~~~~       & 0.0166         & ~~~~~$6.52\times 10^{-8}$~~~~~ \\ \hline
  ~$P$ wave~     & ~~~~~$1.35\times 10^{-6}$~~~~~&$3.90\times 10^{-7}$~ & ~~~~~$4.72\times 10^{-8}$~~~~~&~$9.88\times 10^{-8}$~     \\ \hline\hline
\end{tabular}}
\end{center}
\end{table}

\subsubsection{$\chi_{_{bJ}}(3P) \to \Upsilon+\gamma$}
The three highly excited states $\chi_{_{bJ}}(3P)$ ($J=0,~1,~2$), have same $E_1$ dominated decay channels, they are $\chi_{_{bJ}}(3P)\rightarrow\gamma\Upsilon(nS)$  ($n=1,~2,~3$), and $\chi_{_{bJ}}(3P)\rightarrow\gamma\Upsilon(mD)$ ($m=1,~2$). Among them, $\chi_{_{bJ}}(3P)\rightarrow\gamma\Upsilon(3S)$ present the maximum decay widths, they are $\{11.4,~ 19.2,~ 20.5\}$ keV. The decay processes with final states being $\Upsilon(1S)$ and $\Upsilon(2S)$ also have significant branching ratios. Therefore, we believe that the decay chains, $\chi_{_{bJ}}(3P)\rightarrow\gamma\Upsilon(nS)$, and $\Upsilon(nS)\to \mu^+ \mu^-$ will play important roles in the search for $\chi_{_{bJ}}(3P)$ states in experiments.

In Tables \ref{bottom3P3S0++1-}, \ref{bottom3P3S1++1-} and \ref{bottom3P3S2++1-}, we show the contributions of different partial waves to decay widths of $\chi_{_{bJ}}(3P)$$\to$$ \gamma\Upsilon(3S)$, where $J=0,~1,~2$. The corresponding non-relativistic results of $\Gamma_{non-rel}$ differ slightly from those shown in the tables, they are $\{10.7,~18.3,~19.3 \}$ keV, and the relativistic effects are $\{6.14\%,~4.69\%,~5.85\%\}$.

\begin{table}[H]
\begin{center}
\caption{Contributions of different partial waves to decay width (\rm{keV}) of $\chi_{_{b0}}(3P)$$\to$$ \gamma\Upsilon(3S)$.}\label{bottom3P3S0++1-}
{\begin{tabular}{|c|c|c|c|c|}
\hline
\diagbox{$0^{++}$}{$1^{-}$}&~whole~              & ~$S'$ wave~                & ~$P'$ wave~              & ~$D'$ wave~           \\ \hline
  ~whole~               & 11.4                     & ~11.2~              & ~$6.82\times 10^{-4}$~    &$2.11\times 10^{-5}$       \\ \hline
~~~~~~~$P$ wave~~~~~~~  & 10.9                      & 10.6           & ~~~~~$8.44\times 10^{-4}$~~~~~&$2.81\times 10^{-5}$       \\ \hline
  ~$S$ wave~&~~~~~$6.56\times 10^{-3}$~~~~~& ~~~~~$6.29\times 10^{-3}$~~~~~&$8.68\times 10^{-6}$&~~~~~$5.11\times 10^{-7 }$~~~~~ \\ \hline
\end{tabular}}
\end{center}
\end{table}
\begin{table}[H]
\begin{center}
\caption{Contributions of different partial waves to decay width (\rm{keV}) of $\chi_{_{b1}}(3P)$$\to$$ \gamma\Upsilon(3S)$.}\label{bottom3P3S1++1-}
{\begin{tabular}{|c|c|c|c|c|}
\hline
\diagbox{$1^{++}$}{$1^{-}$}
                       &~whole~                  & ~$S'$ wave~             & ~$P'$ wave~                & ~$D'$ wave~             \\ \hline
  ~whole~              & 19.2                       & ~18.2~           & ~$9.83\times 10^{-3}$~     &~$9.46\times 10^{-6}$~       \\ \hline
~~~~~~~$P$ wave~~~~~~~ & 18.9                        & 18.2         & ~~~~~$4.91\times 10^{-3}$~~~~~&~$4.06\times 10^{-6}$~       \\ \hline
  ~$D$ wave~&~~~~~$1.07\times 10^{-3}$~~~~~&~~~~~$4.20\times 10^{-6}$~~~~~&$9.08\times 10^{-4}$~&~~~~~$1.26\times 10^{-6 }$~~~~~  \\ \hline
\end{tabular}}
\end{center}
\end{table}
\begin{table}[H]
\begin{center}
\caption{Contributions of different partial waves to decay width (\rm{keV}) of $\chi_{_{b2}}(3P)$$\to$$
\gamma\Upsilon(3S)$.}\label{bottom3P3S2++1-}
{\begin{tabular}{|c|c|c|c|c|}
\hline
\diagbox{$2^{++}$}{$1^{-}$}
                           &~whole~                & ~$S'$ wave~             & ~$P'$ wave~             & ~$D'$ wave~            \\ \hline
  ~whole~                   & 20.5                    & ~19.2~                & ~0.0187~            &~$3.33\times 10^{-6}$~     \\ \hline
  ~$P$ wave~                & 19.1                    & ~18.8~     & ~~~~~~$1.58\times 10^{-3}$~~~~~~&$1.50\times 10^{-7}$~     \\ \hline
  ~$D$ wave~               & 0.0129       & ~~~~~~$2.29\times 10^{-5}$~~~~~~   & 0.0119        &~~~~~~$3.94\times 10^{-7}$~~~~~~\\ \hline
~~~~$F$ wave~~~~&~~~~~~$5.29\times 10^{-6}$~~~~~~&$3.33\times 10^{-11}$   &$9.00\times 10^{-8}$      &$4.20\times 10^{-6}$      \\ \hline
\end{tabular}}
\end{center}
\end{table}

We note that, the undiscovered $\Upsilon(2D)$ has significant branching ratios in the decays of $\chi_{_{b0}}(3P) \to \Upsilon(2D)+\gamma$ and $\chi_{_{b1}}(3P) \to \Upsilon(2D)+\gamma$.
Similar to the $\Upsilon(1D)$, $\Upsilon(2D)$ can be also directly produced in $B$ Factory through $e^+ e^-$ annihilation, to search for $\Upsilon(2D)$, the decay chain $\Upsilon(2D)\to \gamma\chi_{_{b0}}(1P,2P)$, $\chi_{_{b0}}(1P,2P)\rightarrow\gamma\Upsilon(1S)$, $\Upsilon(1S)\to \mu^+ \mu^-$ and $\Upsilon(2D)\to \mu^+ \mu^-$ may play important roles.

\section{Summary}
Using the Bethe-Salpeter method, we study the EM transitions between heavy quarkonia,
$\chi_{_{cJ}}\to\gamma+\psi$ and $\chi_{_{bJ}}\to\gamma+\Upsilon$. Emphasis is put on the relativistic effects. Based on the $J^{PC}$ of a particle, we found that its wave function is not composed of pure wave, but contains multiple partial waves. For example, charmonia $\chi_{_{c0}}$, $\chi_{_{c1}}$, $\chi_{_{c2}}$, and $\psi$, and their corresponding bottomonia partners, are $S$$-$$P$, $P$$-$$D$, $P$$-$$D$$-$$F$, and $S$$-$$P$$-$$D$ mixing states, separately. We found that the main wave of each particle provides the non-relativistic contribution, while the other secondary waves give the relativistic corrections. Our results show that for charmonia, relativistic corrections, especially for highly excited states, are relatively important. In the cases of bottomina, the contributions of relativistic corrections are small.

For the two candidates of $\chi_{_{c0}}(2P)$, $\chi_{_{c0}}(3860)$ and $\chi_{_{c0}}(3915)$, their EM decay behavior is completely different, and it is possible to determine which of them is the charmonium $\chi_{_{c0}}(2P)$ based on their electromagnetic decays. Whether $\chi_{_{c1}}(3872)$ is the conventional $\chi_{_{c1}}(2P)$ remains an open question. The calculated decay width of $\chi_{_{c1}}(3872)$$\rightarrow$$\gamma\psi(2S)$ is consistent with data, but the one of $\chi_{_{c1}}(3872)$$\rightarrow$$\gamma\psi(1S)$ is much larger than data. To confirm $\chi_{_{c1}}(3872)$ as a conventional charmonium, the channel $\chi_{_{c1}}(3872)$$\rightarrow$$\gamma\psi(1D)$ can serve as a comparative decay channel. The undiscovered $\chi_{_{c2}}(1F)$ can be found through decay $\chi_{_{c2}}(1F)$$\rightarrow$$\gamma\psi(1D)$, as it has a overwhelming decay width.

For bottomonia, $\chi_{_{b0}}(2P)\rightarrow\gamma\Upsilon(1D)$ and $\chi_{_{bJ}}(3P)\rightarrow\gamma\Upsilon(2D)$ ($J=0,1$) can serve as candidates for discovery of $\Upsilon(1D)$ and $\Upsilon(2D)$, respectively. To search for $\chi_{_{bJ}}(3P)$ $(J=0,1,2)$, the most competitive is the decay $\chi_{_{bJ}}(3P)\rightarrow\gamma\Upsilon(3S)$ followed by $\Upsilon(3S)\to \mu^+ \mu^-$. The best way to find $\chi_{_{b2}}(1F)$ is to search for the decay of $\chi_{_{b2}}(1F)\rightarrow\gamma\Upsilon(1D)$.

{\bf Acknowledgments}
This work was supported in part by the National Natural Science Foundation of China (NSFC) under the Grants Nos. 12075073 and 12375085, the Natural Science Foundation of Hebei province under the Grant No. A2021201009, Post-graduate's Innovation Fund Project of Hebei University under the Grant No. HBU2022BS002.


\begin{thebibliography}{99}

\bibitem{Eichten:1974af}E. Eichten \emph{et al}., Phys. Rev. Lett. \textbf{34}, 369 (1975).
\bibitem{article}W. R. Innes \emph{et al}., Phys. Rev. Lett. \textbf{33}, 1240 (1977).
\bibitem{Gaiser:1985ix}J. E. Gaiser \emph{et al}., (Crystal Ball Collaboration), Phys. Rev. D \textbf{34}, 711 (1986).

\bibitem{BESIII2010-2}M. Ablikim \emph{et al}., (BESIII Collaboration), Phys. Rev. Lett. \textbf{104}, 132002 (2010).
\bibitem{BESIII2017gcu}M. Ablikim \emph{et al}., (BESIII Collaboration), Phys. Rev. D \textbf{96}, 032001 (2017).
\bibitem{BESIII2019-1}M. Ablikim \emph{et al}., (BESIII Collaboration), Phys. Rev. Lett. \textbf{122}, 202001 (2019).

\bibitem{Belle:2003nnu}S. K. Choi \emph{et al}., (Belle Collaboration), Phys. Rev. Lett. \textbf{91} 262001 (2003).
\bibitem{Belle:2010}C.-P. Shen \emph{et al}., (Belle Collaboration), Phys. Rev. Lett. \textbf{104} 112004 (2010).
\bibitem{Belle:2012}R. Mizuk \emph{et al}., (Belle Collaboration), Phys. Rev. Lett. \textbf{109} 232002 (2012).

\bibitem{BaBar1}B. Aubert \emph{et al}., (BaBar Collaboration), Phys. Rev. D. \textbf{81} 092003 (2010).
\bibitem{BaBar2}B. Aubert \emph{et al}., (BaBar Collaboration), Phys. Rev. Lett. \textbf{102} 132001 (2009).
\bibitem{BaBar3}B. Aubert \emph{et al}., (BaBar Collaboration), Phys. Rev. D. \textbf{76} 031102 (2007).

\bibitem{CDF2}T. Aaltonen \emph{et al}., (CDF Collaboration), Phys. Rev. Lett. \textbf{103} 152001 (2009).
\bibitem{CDF:2006ocq}A. Abulencia \emph{et al}., (CDF Collaboration), Phys. Rev. Lett. \textbf{98} 132002 (2007).
\bibitem{CDF3}D. Acosta \emph{et al}., (CDF Collaboration), Phys. Rev. Lett. \textbf{93} 072001 (2004).

\bibitem{D01}V.M. Abazov \emph{et al}., (D0 Collaboration), Phys. Rev. D \textbf{89} 012004 (2014).
\bibitem{D02}V.M. Abazov \emph{et al}., (D0 Collaboration), Phys. Rev. Lett. \textbf{93} 162002 (2004).

\bibitem{BaBar:2006fjg}B. Aubert \emph{et al}., (BaBar Collaboration), Phys. Rev. D. \textbf{74} 071101 (2006).

\bibitem{Belle1}A. Bondar \emph{et al}., (Belle Collaboration), Phys. Rev. Lett. \textbf{108} 122001 (2012).
\bibitem{Belle2}I. Adachi \emph{et al}., (Belle Collaboration), Phys. Rev. Lett. \textbf{108} 032001 (2012).

\bibitem{CLEO:2010xuh}M. Kornicer \emph{et al}., (CLEO Collabration), Phys. Rev. D \textbf{83} (2011).
\bibitem{CLEO2008pkg}D. M. Asner \emph{et al}., (CLEO Collabration), Phys. Rev. D \textbf{78}, 091103 (2008).

\bibitem{Barnes_2005}T. Barnes, S. Godfrey, and E.S. Swanson, Phys. Rev. D \textbf{72} 054026 (2005).
\bibitem{Duan2020tsx}M.-X. Duan, S.-Q. Luo, X. Liu, T. Matsuki, Phys. Rev. D \textbf{101}, 054029 (2020).
\bibitem{Li:2009zu}B.-Q. Li, and K.-T. Chao, Phys. Rev. D \textbf{79}, 094004 (2009).
\bibitem{Segovia:2018qzb}J. Segovia,  S. Steinbei{\ss}er, A. Vairo, Phys. Rev. D \textbf{99} 7, 074011 (2019).
\bibitem{Ebert:2002pp}D. Ebert, R. N. Faustov, and V. O. Galkin, Phys. Rev. D \textbf{67}, 014027 (2003).
\bibitem{Anwar:2018yqm}M. N. Anwar, Y. Lu, B.-S. Zou, Phys. Rev. D \textbf{99}, 094005 (2019).
\bibitem{Brambilla2020ojz}N. Brambilla, H. S. Chung, A. Vairo, Phys. Rev. Lett. \textbf{126}, 082003 (2021).
\bibitem{Baranov2015yea}S. P. Baranov, A. V. Lipatov, and N. P. Zotov, Phys. Rev. D \textbf{93}, 094012 (2016).
\bibitem{SPDMIXING}F.-K. Guo, C. Hanhart, Ulf-G. Mei{\ss}ner, Q. Wang, Q. Zhao, Phys. Lett. B \textbf{725}, 127 (2013).
\bibitem{guo}F.-K. Guo, Ulf-G. Mei{\ss}ner, Phys. Rev. D \textbf{86}, 091501 (2012).
\bibitem{Wang:2010ej}T.-H. Wang, and G.-L. Wang, Phys.Lett.B \textbf{697} 233-237 (2011).
\bibitem{Wang:2022dfd}Z.-H. Wang, G.-L. Wang, Phys.Rev.D \textbf{106}, 054037 (2022).
\bibitem{Chang:2010kj}C.-H. Chang, and G.-L. Wang, Sci. China Phys. Mech. Astron. \textbf{53}, 2005 (2010).

\bibitem{Brambilla2019esw}N. Brambilla, S. Eidelman, C. Hanhart, A. Nefediev, C.-P. Shen, C.E. Thomas, A. Vairo, C.-Z. Yuan, Phys. Rep. \textbf{873}, 1
    (2020).
\bibitem{zsl}H.-X. Chen, W. Chen, X. Liu, Y.-R. Liu, S.-L. Zhu, Rept. Prog. Phys. \textbf{80} 076201 (2017).
\bibitem{Eichten2007qx}E. Eichten, S. Godfrey, H. Mahlke, J.L. Rosner, Rev. Mod. Phys. \textbf{80}, 1161 (2008).
\bibitem{Brambilla2010cs}N. Brambilla \emph{et al}., Eur. Phys. J. C \textbf{71}, 1534 (2011).
\bibitem{Chen2016qju}H.-X. Chen, W. Chen, X. Liu, and S.-L. Zhu, Phys. Rep. \textbf{639}, 1 (2016).

\bibitem{PDG}R. L. Workman \emph{et al}., (Patical Data Group), Prog. Theor. Exp. Phys. \textbf{2022}, 083C01 (2022).

\bibitem{39151}S.-K. Choi \emph{et al}., (Belle Collaboration), Phys. Rev. Lett. \textbf{94} 182002 (2005).
\bibitem{3915}S. Uehara \emph{et al}., (Belle Collaboration), Phys. Rev. Lett. \textbf{104} 092001 (2010).

\bibitem{3860}K. Chilikin \emph{et al}., (Belle Collaboration), Phys. Rev. D \textbf{95} 112003 (2017).

\bibitem{vsquare}G.-L. Wang, T.-F. Feng, X.-G. Wu, Phys. Rev. D \textbf{101}, 116011 (2020).

\bibitem{Salpeter:1951sz}E. E. Salpeter, H. A. Bethe Phys. Rev. \textbf{84}, 1232 (1951).

\bibitem{ret8}E. E. Salpeter, Phys. Rev. \textbf{87}, 328 (1952).

\bibitem{ret1-}G.-L. Wang, Phys. Lett. B \textbf{633}, 492 (2006).

\bibitem{Li:2022qhg}W. Li, S.-Y. Pei, T. Wang, Y.-L. Wang, T.-F. Feng, G.-L. Wang, Phys. Rev. D \textbf{107}, 113002 (2023).

\bibitem{gauge}C.-H. Chang, J.-K. Chen, G.-L. Wang, Commun. Theor. Phys. \textbf{46}, 467 (2006).

\bibitem{Liu:2022kvs}T.-T. Liu,  S.-Y. Pei, W. Li, M. Han, G.-L. Wang, Eur.Phys.J.C \textbf{82} 8, 737 (2022).
\bibitem{ret13}X.-H. Wang, Y. Jiang, T. Wang, X.-Z. Tan, G. Li, G.-L. Wang, Eur. Phys. J. C \textbf{79}, 997 (2019).

\bibitem{Wang:2022cxy}G.-L. Wang, T. Wang, Q. Li, C.-H. Zhang, JHEP, \textbf{05}, 006 (2022).

\bibitem{ret01+}G.-L. Wang, Phys. Lett. B \textbf{650}, 15 (2007).

\bibitem{ret2+}G.-L. Wang, Phys. Lett. B \textbf{674}, 172 (2009).

\bibitem{Eichten:1978tg}E. Eichten, K. Gottfried, T. Kinoshita, K. D. Lane, T.-M. Yan, Phys. Rev. D \textbf{21} 313 (1980).

\bibitem{laermann}E. Laermann, F. Langhammer, I. Schmitt, P.M. Zerwas, Phys. Lett. B {\bf173}, 437 (1986).

\bibitem{Huang:1996bk}Han-Wen Huang, Cong-Feng Qiao, Kuang-Ta Chao, Phys. Rev. D \textbf{54}, 2123 (1996).

\bibitem{Radford:2007vd}S. F. Radford, and W. W. Repko, Phys. Rev. D \textbf{75}, 074031 (2007).
\bibitem{Deng:2016stx}W. J. Deng, H. Liu, L. C. Gui, and X. H. Zhong, Phys. Rev. D, \textbf{95}, 034026 (2016).

\bibitem{Godfrey:2015dia}S. Godfrey, and K. Moats, Phys. Rev. D \textbf{92}, 054034 (2015).
\bibitem{Segovia:2016xqb}J. Segovia, P.G. Ortega, D.R. Entem, F. Fernandez, Phys. Rev. D \textbf{93}, 074027 (2016).
\bibitem{Barducci:2016wze}A. Barducci, R. Giachetti, E. Sorace, Phys. Rev. D \textbf{95}, 054022 (2017).
\bibitem{Deng:2016ktl}W.-J. Deng, H. Liu, L.-C. Gui, and X.-H. Zhong, Phys. Rev. D \textbf{95}, 074002 (2017).
\bibitem{Ananyev:2020uve}V. Ananyev, I. Danilkin, and M. Vanderhaeghen, Phys. Rev. D \textbf{102}, 096019 (2020).

\bibitem{ret9}C. S. Kim, and G.-L. Wang, Phys. Lett. B \textbf{584}, 285(2004).
\end{thebibliography}

 \end{document}